\documentstyle[12pt,axodraw]{article}
\newcommand{\be}{\begin{equation}}
\newcommand{\ee}{\end{equation}}
\newcommand{\bey}{\begin{eqnarray}}
\newcommand{\pslash}{\not{\hbox{\kern-2.3pt $p$}}}
\newcommand{\pdslash}{\not{\hbox{\kern-2pt $\partial$}}}
\newcommand{\bfx}{{\bf x}}
\newcommand{\bfw}{{\bf w}}
\newcommand{\bfy}{{\bf y}}

\newcommand{\barPsi}{{\overline\Psi}}
\newcommand{\barJ}{{\overline{J}}}

\newcommand{\bA}{{\bf A}}
\newcommand{\bB}{{\bf B}}
\newcommand{\bp}{{\bf p}}
\newcommand{\bq}{{\bf q}}
\newcommand{\bk}{{\bf k}}
\newcommand{\beq}{{\begin{equation}}}
\newcommand{\eeq}{{\end{equation}}}
\newcommand{\bea}{{\begin{array}}}
\newcommand{\eea}{{\end{array}}}
\newcommand{\ri}{{\rm i}}

\newcommand{\pd}{\partial}

\newcommand{\bv}{{\bf v}}
\newcommand{\bfn}{{\mathbf{\nabla}}}
\newcommand{\eey}{\end{eqnarray}}

\begin{document}

\begin{titlepage}
\vskip 2cm
\begin{center}
{\Large\bf Path-integral quantization of Galilean Fermi fields}
\footnote{{\tt montigny@phys.ualberta.ca,}
 {\tt khanna@phys.ualberta.ca,}  {\tt fsarajov@phys.ualberta.ca} }
\vskip 3cm
{\bf
M. de Montigny$^{a,b}$, F.C. Khanna$^{a,c}$, F.M. Saradzhev$^{a}$ \\}
\vskip 5pt
{\sl $^a$Theoretical Physics Institute, University of Alberta,\\
 Edmonton, Alberta, Canada T6G 2J1\\}
\vskip 2pt
{\sl $^b$Campus Saint-Jean, University of Alberta, \\
 Edmonton, Alberta, Canada T6C 4G9\\}
\vskip 2pt
{\sl $^c$TRIUMF, 4004, Westbrook Mall,\\
Vancouver, British Columbia, Canada  V6T 2A3\\}
\vskip 2pt

\end{center}
\vskip .5cm
\rm

\begin{abstract}
The Galilei-covariant fermionic field theories are quantized by using the
 path-integral method and five-dimensional Lorentz-like covariant
 expressions of non-relativistic field equations. Firstly we
 review the five-dimensional approach to the Galilean Dirac
 equation, which leads to the L\'evy-Leblond equations, and
 define the Galilean generating functional and Green's
 functions for positive- and negative-energy/mass solutions.
 Then, as an example of interactions, we consider the quartic
 self-interacting potential $\lambda(\barPsi\Psi)^2$, and we
 derive expressions for the 2- and 4-point Green's functions.
 Our results are compatible with those found in the literature
 on non-relativistic many-body systems.
 The extended manifold allows for compact expressions of the
 contributions in $(3+1)$ space-time. This is particularly
 apparent when we represent the results with diagrams in the
 extended $(4+1)$ manifold, since they usually encompass more diagrams in
 Galilean $(3+1)$ space-time.
\end{abstract}

\end{titlepage}

\setcounter{footnote}{0} \setcounter{page}{1} \setcounter{section}{0} %
\setcounter{subsection}{0} \setcounter{subsubsection}{0}

%%%%%%%%%%%%%%%%%%%%%%%%%%%%%%%%%%%%%%%%%%%%%%%%%%%%%%%%%%%%%%%%%%%%%%%%%%%%%%%%%%%

%%%        INTRODUCTION

%%%%%%%%%%%%%%%%%%%%%%%%%%%%%%%%%%%%%%%%%%%%%%%%%%%%%%%%%%%%%%%%%%%%%%%%%%%%%%%

\section{Introduction}

Although its original successes lie in particle physics,
quantum field theory has since then reached a much wider range of applications.
Indeed, concepts such as perturbation methods and Feynman diagrams,
renormalization group procedures, spontaneous symmetry breaking, etc.
(both at zero and finite temperature) have been interchangeably utilized
by physicists working in particle physics as well as in condensed matter
physics and statistical physics \cite{qftcmp,fetter}. More modern field
theories, such as conformal field theory, are applied in statistical
physics, string theory, etc.

However, a fundamental difference between particle physics and condensed
matter (or statistical) physics is that the latter involves the
non-relativistic regime, whereas high-energy physics involves relativistic
kinematics. In fact, field theoretical models typically are constructed by
taking into account various symmetries, such as Poincar\'e space-time
invariance. Recent achievements, such as Fermi condensates with ultra-cold
potassium-40 atoms \cite{jin},suggest that analogous procedures should be
devised for Galilean-invariant systems. Recent interest in the Galilean
symmetry (particularly in the plane) is due to its applications to Hall
efffect, anyons, Chern-Simons term, non-commutative geometry, etc \cite{luk}.

This article is an extension to Fermi fields of a recent work where we have
performed the path-integral quantisation of Galilean-invariant scalar fields
\cite{abreu}. It belongs to a series of papers whose general underlying
program consists in using a formulation of Galilean covariance based on a
relativistic framework in one higher dimension, which makes
non-relativistic field theories similar to Lorentz-covariant
theories \cite{ap1999}-\cite{fluidspaper}. In these articles, the
extended manifold approach follows the lines of earlier investigations
 \cite{takahashi,omote}.
Similar approaches have been ubiquitous in physics \cite{others,kapuscik}.
Recently, it has been used in the study of fluid dynamics
\cite{jackiw,horvathy}.

The occurence of the $2+1$ Galilean group
 was observed \cite{susskind}
 as the transverse motion to the direction of the infinite
 momentum frame, now better known as the light-cone
 frame, in a study of the perturbative behaviour in the limit
 of strong interaction processes. This has been suggested
 previously \cite{weinberg}. Later, this perspective
 was taken up in conjecturing an equivalence between
 eleven-dimensional M-theory and the $N=\infty$-limit of the supersymmetric
 matrix quantum mechanics which describes D0 branes \cite{mtheory} (These
 authors actually consider the super-Galilei group, which admits 32 real
 super-generators.)

Let us review the formalism briefly for our purposes.
The algorithm henceforth consists in building action
functionals by enforcing Lorentz covariance, as it is usually done
with relativistic theories, except that Galilean kinematics
is based on the so-called {\it Galilean five-vectors} $(\bfx,x^4,x^5)$.
These vectors transform under Galilean boosts as
\bey
\bfx'&=&\bfx-{\bf{\beta}}  x^4,\nonumber\\
x^{4'}&=&x^4,\nonumber\\
x^{5'}&=&x^5-{\bf{\beta}} \cdot\bfx +\frac{1}{2}{|{\bf{\beta}}|}^{2} x^4,\label{strans}
\eey
where ${\bf{\beta}} \equiv {\bv}/\bar{c}$, while ${\bv}$  is the relative velocity between the two reference frames,
and $\bar{c}$ is a parameter with the dimensions of velocity, which will be
specified below.

Altogether the kinematical transformations, which also include
rotations and translations, form a fifteen-dimensional Lie algebra. This
may be seen as the Poincar\'e algebra in $(4+1)$ space-time. Eleven of these
fifteen generators form the extended Galilei group, where the central-extension
parameter (the non-relativistic mass) is inherited from the generator of $x^5$
translations. The transformation (\ref{strans}) for $x^{\prime\ 5}$ has
occurred in various contexts \cite{abreu}-\cite{kapuscik}. In quantum
mechanics, it is associated with the wave function's phase which enforces
invariance of the Schr\"odinger equation under Galilean transformations.
Furhermore, it leads to a superselection rule for mass conservation
 in Galilean covariant theories. Indeed,
unlike the relativistic theories, new massive particles cannot be created
in a Galilean framework. This would preclude Yukawa couplings except for
 massless particles with coupling to two massive particles. For massive
 particles, only the 4-particle coupling is allowed.

%In the classical context, a lucid physical interpretation of $x^5$ as a control
% parameter which relates the two velocities (ongoing and returning)
% of a non-relativistic signal has been given by Kapu\'scik (1986)
% using a description reminiscent of customary
% pedagogical presentations of Einstein's relativity (in terms of signals, clocks,
% rods and mirrors) \cite{kapuscik}. Indeed, for a given signal, the sum of
% the ongoing and returning velocities is the same for different observers,
% but the control parameter helps to keep track of different signals.
% In the Einsteinian case, this question does not occur since only one
% fundamental constant, the velocity of light, is sufficient since it is
% the same in all reference frames.

The invariant scalar product is defined as
\[
A\cdot B=A_4B_5 + A_5B_4 - \bA\cdot\bB,
\]
with the {\em Galilean metric}~:
\bey
g_{\mu\nu}=\left( \begin{array}{ccc} -{\mathbf 1}_{3\times 3}
& 0 & 0 \\ 0 & 0 & 1 \\ 0 & 1 & 0\end{array} \right).\label{galileanmetric}
\eey
This suggests that the non-relativistic time is a light-cone
parameter of the Lorentz invariant theory on a
 manifold containing one additional space-like
dimension \cite{omote,zwiebach}. Once we have constructed a Galilean covariant
action functional, an appropriate embedding of the Galilean space-time
into ${{\cal G}}_{\rm (4+1)}$ may be defined as
\[
(\bfx, t)\hookrightarrow x^\mu=(x^1,\cdots, x^5)\equiv \left (\bfx, \bar{c}t, \frac{s}{\bar{c}}\right ).
\]

The five-momentum
\bey
p_\mu=\ri\pd_\mu=
 \left (\ri\bfn,\frac{\ri\pd_t}{\bar{c}},\ri\bar{c} \pd_s\right )=\left (\bp, {\frac{E}{\bar{c}}} , m\bar{c} \right ),
\label{fivemomentum} \eey
where $p^4=p_5=m\bar{c}$ and $p^5=p_4=\frac{E}{\bar{c}}$, suggests that the additional
coordinate $x^5=\frac{s}{\bar{c}}$ is canonically conjugated to $m\bar{c}$.
From the relation $\pd_s=-\ri m$, the phase factor of the wavefunction follows:
\bey
\Psi (x)\equiv e^{-\ri m s}\psi (\bfx,t),\label{embeddingcplxf}
\eey
which projects the fields from ${\cal G}_{(4+1)}$ to $(3+1)$-dimensions.
A different definition of dimensional reduction would lead to a Lorentz-covariant theory in
$(3+1)$-dimensions \cite{omote,others}.

Note that it is also possible to define
\[
\Psi (x)\equiv e^{-\ri m s}\psi_+ (\bfx,t)+ e^{+\ri m s}\psi_-  (\bfx,t),
\]
where $\psi_\pm (\bfx,t)$ represent the positive- or negative-energy solutions,
which makes evident the possibility of negative energy solutions
\cite{horzela}. This comes from the quadratic condition $(\pd_s)^2=-m^2$, and it is compatible
with the embedding defined in Eq. (\ref{embeddingcplxf}) since
additional terms with negative mass and negative energy can be included.
Such a description is allowed by the symmetry $(\bar{c}t,\frac{s}{\bar{c}})\rightarrow (-\bar{c}t,-\frac{s}{\bar{c}})$
\cite{santos2}.

In $(4+1)$-dimensional Galilean theories, $p_{\mu} p^{\mu}=2mE-{\bp}^2$ is an
invariant, and the dynamics of Galilean covariant fields must be consistent with it.
Let us take
\[
p_{\mu} p^{\mu}=k^2,
\]
where $k$ is a real constant that defines the invariant quantity.
 It leads to $2mE-{\bf p}^2=k^2$, which
 is analogous to $E^2-p^2c^2$, the invariant
 for Lorentz covariant fields that is equal to
$m^2c^4$, thus defining $m$ as the invariant quantity.
This implies the dispersion relation~:
\bey
 E=\frac 1{2m}|\bp|^2 +\frac 1{2m} k^2.\label{disprel}
\eey
Introducing a velocity parameter
\[
\bar{c}=\frac{k}{{\sqrt{2}}m},
\]
we cast the dispersion relation into the familiar form for non-relativistic energy
with $E=\frac 1{2m}|\bp|^2 + m {\bar{c}}^2$. Note that this equation, as well
 as the invariant  $2mE-{\bf p}^2$, is invariant under the changes
 \[
m\longrightarrow -m,\qquad E\longrightarrow -E,
\]
so that in the $(4+1)$-manifold, one must reverse both
 $m$ and $E$, but not each one independently. Henceforth,
 we will have to ensure that this is satisfied when we
 split the positive- and negative-energy/mass solutions.

The constant $k$ is the Galilean analogue of
the Lorentzian rest mass. Since $k$ can be absorbed within the
energy $E$, its value is usually considered to be of no physical
importance and taken to be zero. However, it may be possible to
relate $k$ to the chemical potential \cite{abreu}.
For Galilean Fermi fields, the dispersion relation (\ref{disprel})
implies that the negative energy solutions are characterized by
negative masses.

The paper is organized as follows. In Section 2, we review the L\'evy-Leblond
 equations by means of the extended-manifold Dirac equation, and the
 positive- and negative-energy/mass solutions, and we
 introduce the Galilean generating functional formalism.
 Appropriate embeddings associated with the virtual sources are
 defined.  In Section 3, we establish the connection between the
 generating functional and the Green's functions for both
 positive- and negative energy/mass solutions. In Section 4, we
 apply this formalism to the self-interacting quartic potential.
 The 2- and 4-point functions are calculated. We distinguished
 between $(3+1)$- and $(4+1)$-manifold diagrams, the latter
 containing, in general, more diagrams in the reduced $(3+1)$
 space-time. Concluding remarks are in Section 5.

%%%%%%%%%%%%%%%%%%%%%%%%%%%%%%%%%%%%%%%%%%%%%%%%%%%%%%%%%%%%%%%%%%%%%%%%%%%%%%%%%%
%
%
%%%%%%%%%%%%%%%%%%%% DIRAC FIELD - FREE Z[J]
%
%
%%%%%%%%%%%%%%%%%%%%%%%%%%%%%%%%%%%%%%%%%%%%%%%%%%%%%%%%%%%%%%%%%%%%%%%%%%%%%%%%%%%

\section{Free Dirac field}
%%%%%%%%%%%%%%%%%%%%%%%%%%%%%%%%%%%%%%%%%%%%%%%%%%%%%%%%%%%%%%%%%%%%%%%
\subsection{Five-dimensional Dirac equation}
%%%%%%%%%%%%%%%%%%%%%%%%%%%%%%%%%%%%%%%%%%%%%%%%%%%%%%%%%%%%%%%%%%%%%

Let us consider a free
Dirac field ${\Psi}(x)$ defined on the five-dimensional
manifold ${{\cal G}}_{\rm (4+1)}$ with Galilean metric,
Eq. (\ref{galileanmetric}). Then a manifestly covariant Lagrangian
 for the Dirac field is given by
\bey {\cal L}_0= \overline{\Psi}(x)(\ri \gamma^\mu
\stackrel{\leftrightarrow}{\partial_\mu}-k)\Psi(x), \label{lagdirac}
\eey
where
 $a\stackrel{\leftrightarrow}{\partial b}\equiv
 \frac 12\left [a\partial b - (\partial a)b\right ]$.
 Both the field and its adjoint are anticommuting.
The matrices $\gamma^{\mu}$ in the extended space-time
are four-dimensional and may be chosen as
\[
\gamma^a=
\left(\begin{array}{cc}
\ri\sigma^a&0\\
0&-\ri\sigma^a
\end{array}\right),\;\;\;
\gamma^{4}=\left(
\begin{array}{cc}
0&0\\
\sqrt{2}&0
\end{array}\right),\;\;\;
\gamma^5=\left(
\begin{array}{cc}
0&\sqrt{2}\\
0&0
\end{array}\right),
\]
${\sigma}^a$, $a=1,2,3$ denoting the $2 \times 2$ Pauli matrices.
They obey the usual anticommutation relations :
\[
\left\{ \gamma^{\mu},\gamma^{\nu}\right\}=2g^{\mu \nu}.
\]

Let us apply the variational principle for the action integral
 with the free Lagrangian of Eq. (\ref{lagdirac}),
\bey
I[\Psi,\barPsi]=\int d^{5}x\ {\cal L}_0[\Psi,\barPsi],
\label{acao}
\eey
where the integral over $x^5$ is interpreted
as $\int dx^5\rightarrow\lim_{l\rightarrow\infty}\frac{1}{l}\int_{-l/2}^{l/2} dx^5$,
 and $l$ is an arbitrary length. Then the Euler-Lagrange equations of motion
 for $\Psi(x)$ and its adjoint $\overline{\Psi}(x)$, respectively, are
\bey \left(\ri
\gamma^{\mu}\partial_{\mu}-k\right)\Psi(x)=0,\qquad\qquad
\overline{\Psi}(x) (\ri
\gamma^\mu\stackrel{\leftarrow}{\partial}_\mu+k)=0,\label{d1} \eey
where $a \stackrel\leftarrow\partial b=(\partial a)b$.
The adjoint field is defined as
\[
\overline{\Psi}(x)=\Psi^\dagger(x)\ \gamma^0,\]
where
\[
\gamma^0=\frac{1}{\sqrt{2}}\left(\gamma^4+\gamma^5\right)=
\left(
\begin{array}{cc}
0 & 1\\
1 & 0
\end{array}
\right).
\]
Its reduction to $(3+1)$-dimensions is defined as
\begin{equation}
\bar{\Psi}(x) = e^{\ri m\bar{c}x^5} \bar{\psi}({\bf x},t).
\label{newpro}
\end{equation}

The first expression in Eq. (\ref{d1}),
using Eqs. (\ref{fivemomentum}) and (\ref{embeddingcplxf}), reduces to
\begin{equation}
\begin{array}{l}
\left( ( {\bf \sigma} \cdot {\bfn}) + k
\right)\psi_{1}(\bfx,t)-\sqrt{2}m\bar{c}\psi_{2}(\bfx,t)=0,\\
{\sqrt{2}}\frac{E}{\bar{c}} \psi_{1}(\bfx,t)+ \left( ({\bf \sigma} \cdot {\bfn}) -
k \right)\psi_{2}(\bfx,t)=0,\end{array} \label{pervoe}
\end{equation}
where
\[
\psi(\bfx,t)=\left (\begin{array}{c}\psi_{1}(\bfx,t)\\
{\psi_{2}(\bfx,t)}\end{array}\right ).
\]
 with $\psi_1(x,t)$ and $\psi_2(x,t)$ being two-component spinors.
 The two equations in Eq. (\ref{pervoe}) are analogous to the Pauli
 equations in the relativistic case.
These Galilean wave equations describe non-relativistic Fermi fields
in $(3+1)$-dimensions. If $k=0$ then
Eq. (\ref{pervoe}) coincides with the L\'{e}vy-Leblond
equations
\cite{levyleblond1967}. The wave equations for the adjoint Fermi fields
have the same form and can be deduced from the second expression in
Eq. (\ref{d1}) together with Eq. (\ref{newpro}).

In analogy with the relativistic theory, we find that the Fourier
components of the Galilean Dirac fields satisfy
$\left(p_{\mu}p^{\mu}-k^2\right){\Psi}(p)=0$ and
$\left(p_{\mu}p^{\mu}-k^2\right){\bar{\Psi}}(p)=0$, which
reduce to the Schr\"odinger wave equations. Then each component of
the $(3+1)$-dimensional non-relativistic Fermi fields obeys the
Schr\"{o}dinger equation :
\[ E{\psi}_{1,2}(p)=
\left(\frac{{\bf{p}}^2}{2m}+\frac{k^2}{2m}\right){\psi}_{1,2}(p).
\]
Thus the Schr\"odinger equation may be obtained either by first reducing
the Dirac equation to the L\'{e}vy-Leblond equations (\ref{pervoe}) with
Eq. (\ref{embeddingcplxf}), or by first reducing
Eq. (\ref{lagdirac}) to the Lagrangian of the Schr\"{o}dinger field.

%%%%%%%%%%%%%%%%%%%%%%%%%%%%%%%%%%%%%%%%%%%%%%%%%%%%%%%%%%%%%%%%%%%%%%%%%%%
\subsection{The positive- and negative-energy solutions and canonical quantization}
%%%%%%%%%%%%%%%%%%%%%%%%%%%%%%%%%%%%%%%%%%%%%%%%%%%%%%%%%%%%%%%%%%%%%%%%%

The Lagrangian, Eq. (\ref{lagdirac}), and the Dirac equations given by
Eq. (\ref{d1}) are invariant with respect to unitary transformations
\[
{\gamma}^{\mu} \to S {\gamma}^{\mu} S^{-1}, \qquad {\Psi} \to S {\Psi},
\]
where $S$ is a  $4 \times 4$-matrix. To construct the positive- and
 negative-energy/mass solutions explicitly, it is convenient to use a representation
in which ${\gamma}^0$ is diagonal. This representation
can be obtained from the one used in the previous section by
performing the unitary transformation above with the choice
\[
S=\frac{1}{\sqrt{2}}
\left(
\begin{array}{cc}
1 & 1\\
1 & -1
\end{array}
\right).
\]
In particular, ${\gamma}^0$ becomes
\begin{equation}
{\gamma}^0=
\left(
\begin{array}{cc}
1 & 0\\
0 & -1
\end{array}
\right).
\label{sgammazero}
\end{equation}
It is important to point out that this form of ${\gamma}^0$ matrix
does not imply chirality. In fact, there is no parity operator in five dimensions,
hence no chirality. Only if we work in even dimensions, in this case six dimensions,
can we find a parity operator, hence a chirality operator. Then the ${\gamma}$-matrices
are $8$-dimensional. The details of this representation will appear elsewhere
 \cite{kobayashi}.

The matrices ${\gamma}^{\mu}$ take the form
\bey
\gamma^a=
\left(\begin{array}{cc}
0 & \ri\sigma^a\\
\ri\sigma^a & 0
\end{array}\right),\;\;\;
\gamma^{4}=\frac{1}{\sqrt{2}} \left(
\begin{array}{cc}
1 & 1 \\
-1 & -1
\end{array}\right),\;\;\;
\gamma^5=\frac{1}{\sqrt{2}} \left(
\begin{array}{cc}
1 & -1\\
1 & -1
\end{array}\right). \label{gsmatrices}
\eey
In what follows, we will use the representation defined by equations
(\ref{sgammazero}) and (\ref{gsmatrices}).

The plane-wave solutions for Eq. (\ref{d1}) are written in the usual form,
\[
{\Psi}^{(r)}(x) = \frac 1{(2\pi)^5}\int\;d^5p\
 \left [ u^{(r)}(p) e^{-ipx} + v^{(r)}(p) e^{ipx}\right ],
\]
where $r=1,2$ and the positive- and negative-energy spinors $u^{(r)}(p)=
u^{(r)}({\bp}, E, m)$, $v^{(r)}(p)=v^{(r)}({\bp}, E, m)$ obey the equations
\be
({\gamma}^{\mu} p_{\mu} -k) u^{(r)}(p) =  0, \qquad\quad
({\gamma}^{\mu} p_{\mu} +k) v^{(r)}(p)  =  0.
\label{vp}
\ee
Taking the Dirac particle in the rest frame, ${\bf p}=0$, we find
\[
{\gamma}^0 u^{(r)}(0)  =  u^{(r)}(0), \qquad\quad
{\gamma}^0 v^{(r)}(0)  =  - v^{(r)}(0),
\]
where
\[
u^{(r)}(0) \equiv u^{(r)}(0, E_k, m),
\]
\[
v^{(r)}(0) \equiv v^{(r)}(0, E_k, m),
\]
and
\[
E_k \equiv \frac{k^2}{2m}.
\]
 The representation with diagonal ${\gamma}^0$ is especially appropriate for
describing particles at rest, the spinors $u^{(r)}(0)$, $v^{(r)}(0)$ being
eigenvectors of ${\gamma}^0$ with eigenvalues $+1$ and $-1$,
respectively.

Let us define
\[
{\xi}^{(1)}(0)=
\left(
\begin{array}{c}
1\\
0\\
\end{array}
\right), \qquad
{\xi}^{(2)}(0)=
\left(
\begin{array}{c}
0\\
1\\
\end{array}
\right),
\]
which satisfy the relation
\[
{\xi}^{{\dagger} (r)}(0) {\xi}^{(s)}(0) = {\delta}_{rs}.
\]
We write the spinors $u^{(r)}(0)$, $v^{(r)}(0)$ as
\[
u^{(r)}(0)=
\left(
\begin{array}{c}
{\xi}^{(r)}(0)\\
0\\
\end{array}
\right), \qquad
v^{(r)}(0)=
\left(
\begin{array}{c}
0\\
{\xi}^{(r)}(0)\\
\end{array}
\right).
\]

In a moving frame, the spinors $u^{(r)}(p)$, $v^{(r)}(p)$
 are expressed as
\be
u^{(r)}(p)  =  d_u ({\gamma}^{\mu} p_{\mu} + k )u^{(r)}(0),
 \qquad\quad
v^{(r)}(p)  =  d_v ({\gamma}^{\mu} p_{\mu} - k )v^{(r)}(0).
\label{dvp}
\ee
These definitions are motivated by $p_{\mu}p^{\mu}=k^2$, so that $({\gamma}^{\mu}
p_{\mu} + k)({\gamma}^{\mu} p_{\mu} - k)= p^2 - k^2=0$ and
Eq. (\ref{vp}) is satisfied.

The coefficients $d_u$, $d_v$ are computed from two conditions. Firstly,
the right- and left-hand sides of Eq. (\ref{dvp}) must
coincide for ${\bf p}=0$, that is, $u^{(r)}({\bf p}=0)=u^{(r)}(0)$ and
$v^{(r)}({\bf p}=0)=v^{(r)}(0)$. Secondly, the orthonormality conditions
\[
\bar{u}^{(r)}(p) u^{(s)}(p)  =  {\delta}_{rs},\qquad\quad
\bar{v}^{(r)}(p) v^{(s)}(p)  =  - {\delta}_{rs},
\]
which can be checked for the ${\bf p}=0$ case, must be valid for
non-zero ${\bf p}$ as well. These two conditions determine
 $d_u$ and $d_v$ as
\[
d_u=-d_v=\frac{1}{2k} {\left( \frac{4E_k}{E+3E_k} \right)}^{1/2}.
\]

The general solution to the Galilean Dirac equations (\ref{d1}) may be
expanded in terms of the plane wave solutions as
\begin{eqnarray}
{\Psi}(x) & = & \frac{1}{(2\pi)^{3/2}} \sum_{r} \int d^3\bp
\Big[ a^{(r)}(\bp) u^{(r)}(p) e^{-ipx} + b^{{\dagger} (r)}(\bp) v^{(r)}(p)
e^{ipx} \Big],
\label{psiexpansion} \\
\bar{\Psi}(x) & = & \frac{1}{(2\pi)^{3/2}} \sum_{r} \int d^3\bp
\Big[ a^{{\dagger} (r)}(\bp) \bar{u}^{(r)}(p) e^{ipx} + b^{(r)}(\bp)
\bar{v}^{(r)}(p)
e^{-ipx} \Big],
\label{barpsiexpansion}
\end{eqnarray}
where $a^{(r)}(\bp)$ ($a^{{\dagger} (r)}(\bp)$) and
 $b^{(r)}(\bp)$ ($b^{{\dagger} (r)}(\bp)$)  are
 destruction (creation) operators of particles and antiparticles,
 respectively. The fields are quantised by assuming that these operators
 obey the anticommutation relations:
\[
\Big\{ a^{(r)}(\bp), a^{{\dagger} (s)}(\bq) \Big\} =
\Big\{ b^{(r)}(\bp), b^{{\dagger} (s)}(\bq) \Big\}=
{\delta}_{rs} {\delta}({\bf p} - {\bf q}).
\]
All other anticommutation relations are zero.

Using the non-relativistic ``momentum-energy-mass'' tensor
\[
T^{{\mu}{\nu}} = \frac{{\partial}{\cal
L}}{{\partial}({\partial}_{\mu}{\Psi})}
{\partial}^{\nu} {\Psi} + {\partial}^{\nu} \bar{\Psi}
\frac{{\partial}{\cal L}}{{\partial}({\partial}_{\mu}\bar{\Psi})}
- {\cal L} g^{{\mu}{\nu}},
\]
we define the five-momentum of the Galilean Dirac field as
\[
P_{\mu} = \int d^3x dx^5 \sqrt{2}\; T_{5{\mu}}
=\frac{i}{\sqrt{2}} \int d^3x dx^5 \Big( \bar{\Psi} {\gamma}^4
{\partial}_{\mu}{\Psi} -{\partial}_{\mu} \bar{\Psi} {\gamma}^4
{\Psi}).
\]
The charge operator is written as
\[
Q=\int d^3x dx^5 \; \sqrt{2}\bar{\Psi} {\gamma}^4 {\Psi}.
\]
Substituting the expansions (\ref{psiexpansion}) and
 (\ref{barpsiexpansion}) into the expressions for
$P_{\mu}$ and $Q$ and performing a normal ordering with respect to the
vacuum state, we get
\[
a(\bk)|0\rangle=b(\bk)|0\rangle=0 \qquad {\rm for} \hspace{3 mm} {\rm  all}
\hspace{3 mm} {\bk}  \hspace{3 mm} {\rm and} \hspace{3 mm} m,
\]
giving, for the five-momentum and the charge operators,
\[
P^{\mu} = \sum_{r} \int d^3\bp \cdot p^{\mu}
\Big[ a^{{\dagger}(r)}(\bp) a^{(r)}(\bp) + b^{{\dagger}(r)}(\bp) b^{(r)}(\bp)
\Big],
\]
and
\[
Q = \sum_{r} \int d^3\bp \cdot
\Big[ a^{{\dagger}(r)}(\bp) a^{(r)}(\bp) - b^{{\dagger}(r)}(\bp) b^{(r)}(\bp)
\Big].
\]
This corroborates the point, mentioned earlier, that
 $a^{\dagger}(\bp)$ and $a(\bp)$ are the creation
and annihilation operators for particles of momentum ${\bf p}$, mass $m$
and charge $+1$, whereas the operators $b^{\dagger}(\bp)$ and $b(\bp)$ correspond
to antiparticles, which differ from the particles only by the sign of
the charge, i.e.  $-1$.

%%%%%%%%%%%%%%%%%%%%%%%%%%%%%%%%%%%%%%%%%%%%%%%%%%%%%%%%%%%%%%%%%%%

\subsection{Galilean generating functional}

%%%%%%%%%%%%%%%%%%%%%%%%%%%%%%%%%%%%%%%%%%%%%%%%%%%%%%%%%%%%%%%%%%%%%

As in the usual path-integral formalism \cite{fieldtheorybooks},
 the Galilean generating functional for the {\em free field}
 is given by the vacuum-to-vacuum transition amplitude
 with anticommuting virtual sources $J\left( x\right)$ and
 $\barJ \left( x\right)$ :
\[
Z_0\left[ J,\barJ\right] =\int {\cal D}{\barPsi} \int {\cal
D}{\Psi} \exp
\left\{ \ri\int d^{5}x\left[ {\cal L}_0\left[ {\Psi} ,{\barPsi} \right]
+ {\barPsi} \left( x\right) J\left( x\right) +\barJ\left(
x\right)
{\Psi} \left( x\right) \right] \right\},
\]
where $\int {\cal D}{\barPsi}$ and $\int {\cal D}{\Psi}$ denote the
functional integrations over ${\barPsi}(x)$ and ${\Psi}(x)$, respectively.
 Here $J(x)$ and $\barJ(x)$ are anticommuting Grassmann virtual sources that we put
equal to zero at the end.

Let us define a new field ${\Psi}^{\prime}$ :
\[
{\Psi}^{\prime} \left( x\right)={\Psi} \left( x\right) - {\Psi}_{J}(x),
\]
where ${\Psi}_{J}(x)$ satisfies the inhomogeneous equation of motion :
\bey \left( \ri {\gamma}^{\mu} {\partial}_{\mu} -k \right)
{\Psi}_{J}(x) = - J(x). \label{inho} \eey
Then we can complete the square within the exponential and rewrite the generating functional as
\[
\begin{array}{rcl}
Z_{0}\left[ J,\barJ\right] & = & \int {\cal D}{\Psi}^{\prime }{\cal
D}{\barPsi}^{\prime }\exp \Big\{ \ri\int d^{5}x\ {\cal L}_{0}\left(
{\Psi}^{\prime },{\barPsi}^{\prime }{}\right ) \\
&  & \qquad\qquad +\frac{\ri}{2} \int
d^{5}x \left( {\barJ} (x) {\Psi}_{J}(x) + {\barPsi}_{J}(x)
J(x) \right) \Big\} .
\end{array}
\]
Here we have changed the integration variables from ${\Psi}$
to ${\Psi}^{\prime }$, for which  the Jacobian is unity. Denoting the
integration over ${\Psi}^{\prime }$ and ${\barPsi}^{\prime }{}$ by
$Z_{0}%
\left[ 0\right]$, we observe that the generating functional becomes
\[
Z_{0}\left[ J,\barJ\right] =Z_{0}\left[ 0\right]\ \exp \left\{
\frac{\ri}{2} \int d^{5}x\  \left( \barJ\left( x\right)
{\Psi}_{J}(x) + {\barPsi}_{J}(x) J\left( x\right) \right) \right\} .
\]

The field ${\Psi}_J(x)$ can be written as
\begin{equation}
{\Psi}_J(x) = - \int d^5y S_1(x-y) J(y),
\label{psij}
\end{equation}
where $S_1(x-y)$ is the free-field Green's function, which satisfies
\begin{equation}
(\ri {\gamma}^{\mu} {\partial}_{\mu} -k)
S_1(x-y)=\tilde{\delta}^5(x-y), \label{sf}
\end{equation}
where we adopt a non-standard definition of the delta function:
\begin{equation}
\tilde{\delta}^5(x-y) \equiv \tilde{\delta}^5(x-y;m)+\tilde{\delta}^5(x-y;-m),
\label{deltafunction}
\end{equation}
with $\tilde{\delta}^5(x-y;\pm m)$ being Dirac delta functions \cite{abreu} in
${{\cal G}}_{\rm (4+1)}$ defined as
\[
\tilde{\delta}^5(x-y;\pm m) = e^{\mp \ri m\bar{c} (x^5-y^5)}
{\delta}^3({\bfx}-
{\bfy}) {\delta}(x^4-y^4).
\]
Taking $S_1(x-y)$ as
\begin{equation}
S_1(x-y)= -\frac{\ri {\gamma}^{\mu} {\partial}_{\mu} + k}{2k}
{\Delta}(x-y), \label{relation}
\end{equation}
and introducing the $\ri {\varepsilon}$ prescription by replacing $k^2$
with
$k^2 - \ri {\varepsilon}$, we bring Eq. (\ref{sf}) into the form
\begin{equation}
\frac{1}{2k} \left( {\partial}_{\mu} {\partial}^{\mu} + k^2 - \ri
{\varepsilon} \right) {\Delta}(x-y) = \tilde{\delta}^5(x-y),
\label{scalarprop}
\end{equation}
which is the equation for the Feynman propagator for a
free Galilean scalar field. In the next section, we will show that
${\Delta}(x-y)$ coincides with the Galilean Feynman propagator up to a
constant factor.

The equation for the adjoint field
$\bar{\Psi}_J(x)$,
\[
\bar{\Psi}_J(x) (\ri
{\gamma}^\mu\stackrel{\leftarrow}{\partial}_\mu+k)= \bar{J}(x),
\]
is solved by
\begin{equation}
\bar{\Psi}_J(x) = - \int d^5y \bar{J}(y) S_2(x-y),
\label{barpsij}
\end{equation}
where $S_2(x-y)$ is a solution of
\[
S_2(x-y) (\ri {\gamma}^\mu\stackrel{\leftarrow}{\partial}_\mu+k)=
-\tilde{\delta}^5(x-y).
\]

The function $S_2(x-y)$ is related to ${\Delta}(x-y)$ by
\[
S_2(x-y)= \frac{1}{2k} {\Delta}(x-y) (\ri
{\gamma}^\mu\stackrel{\leftarrow}{\partial}_\mu-k),
\]
so that
\[
S_2(y-x)=S_1(x-y).
\]

For the external source $J$, we factor out the coordinate
$x^5$ as follows :
\[
J\left( x\right) = e^{-\ri m\bar{c} x^5} j_{+} ({\bf x,} x^4)
+ e^{\ri m\bar{c} x^5} j_{-} ({\bf x,} x^4),
\]
and similarly for $\barJ(x)$. This factorization is motivated by
the definition of the fields
${\Psi}_J(x)$ and $\bar{\Psi}_J(x)$ in
Eqs. (\ref{psij}) and (\ref{barpsij}), respectively,
 so that we have, for instance :
\[
\begin{array}{rcl}
(\ri \gamma^\mu \partial_\mu -k )\Psi_J(x)& = &
 - \int d^5y\; \tilde{\delta}^5(x-y) J(y),\\
 & = & \lim_{l\rightarrow\infty} \frac{1}{l} \int_{-l/2}^{l/2} dy^5\;
\left[ e^{-\ri m\bar{c}(x^5-y^5)} + e^{\ri m\bar{c}(x^5-y^5)} \right] \times\\
& & \qquad\times \left[ e^{-\ri m\bar{c}y^5} j_{+}({\bf x},x^4) +
e^{\ri m\bar{c}y^5} j_{-}({\bf x},x^4) \right]\\
& = & - J(x),\end{array}
\]
in agreement with Eq. (\ref{inho}). Note that in the second line,
 we have eliminated $y$ by integration and by using the definition
 of the delta function given in Eq. (\ref{deltafunction}).

With the fields ${\Psi}_{J}(x)$ and $\bar{\Psi}_{J}(x)$ defined by
Eqs. (\ref{psij}) and (\ref{barpsij}) respectively, the generating
functional takes the form
\begin{equation}
Z_{0}\left[ J,\barJ\right] =Z_{0}\left[ 0\right]\ \exp \left\{
-{\ri} \int d^{5}x\ d^5y \barJ\left( x\right) S_1(x-y)
 J\left(y\right) \right\}.  \label{z0final}
\end{equation}
This is the generating
functional of the Green's function that characterizes the Dirac
field. Note that $Z_0\left[ J,\bar{J} \right]$ is written in terms
of the propagator $S_1(x-y)$, which includes both particle and
antiparticle contributions.

%%%%%%%%%%%%%%%%%%%%%%%%%%%%%%%%%%%%%%%%%%%%%%%%%%%%%%%%%%%%%%%%%%%%

\section{Green's functions for particles and antiparticles}

%%%%%%%%%%%%%%%%%%%%%%%%%%%%%%%%%%%%%%%%%%%%%%%%%%%%%%%%%%%%%%%%%%%%%

Let us now turn to some properties of the Galilean propagators
${\Delta}(x-y)$ and $S_1(x-y)$. The Fourier transforms of these
propagators are defined by expressions similar to the mass-shell
condition:
\begin{eqnarray}
{\Delta}(x-y) & = & \frac{1}{(2\pi)^5} \int d^5p\; \bar{\Delta}(p)
e^{-\ri p (x-y)}\; 2\pi\; \left[ {\delta}(p^4- m\bar{c})
+ {\delta}(p^4+ m\bar{c}) \right],
\label{deltatransform}\\
S_1(x-y) & = & \frac{1}{(2\pi)^5} \int d^5p\;
\bar{S}_1(p)
e^{-\ri p (x-y)} \; 2\pi\; \left[ {\delta}(p^4-m\bar{c})
+ {\delta}(p^4+ m\bar{c}) \right].
\label{stransform}
\end{eqnarray}
By substituting Eq. (\ref{deltatransform}) into Eq. (\ref{scalarprop}) and
using
\[
\tilde{\delta}^5(x-y) =  \frac{1}{(2\pi)^5} \int d^5p
e^{-\ri p (x-y)} \;2\pi\; \left[ {\delta}(p^4- m\bar{c})
+ {\delta}(p^4+ m\bar{c}) \right],
\]
where the term between brackets is reminiscent of our non-standard
 definition of delta function in $(4+1)$ dimensions, we find
\[
\bar{\Delta}(p)= \frac{-2k}{p_{\mu}p^{\mu} - k^2
+\ri{\varepsilon}},
\]
so that we rewrite ${\Delta}(x-y)$ as
\[
{\Delta}(x-y)=\frac{\sqrt{2}}{(2{\pi})^4} \int d^3\bp \int dp^5\;
e^{\ri {\bf p}({\bf x}-{\bf y}) -\ri p^5 (x^4-y^4)}
\left[ \frac{e^{\ri m\bar{c}(x^5-y^5)}}{p^5+\frac{E}{\bar{c}} -\ri
\frac{\varepsilon}{2m\bar{c}}} - \frac{e^{-\ri m\bar{c}(x^5-y^5)}}{p^5-\frac{E}{\bar{c}}
+\ri \frac{\varepsilon}{2m\bar{c}}} \right].
\]
Integrating over $p^5$ with the change of variable $p^5 \to p^5 +
\frac{E}{\bar{c}}$ in the first integral, and $p^5 \to p^5 - \frac{E}{\bar{c}}$ in
the second one, and by using the following representation of the step
function :
\[
{\theta}(\tau)= \lim_{{\varepsilon} \to 0^{+}}
\frac{-1}{2{\pi}\ri} \int_{-\infty}^{\infty} d{\omega}\;
\frac{e^{-\ri{\omega}{\tau}}}{{\omega}+\ri{\varepsilon}},
\]
we obtain
\[
{\Delta}(x-y)=\sqrt{2}\; {\Delta}_F(x-y),
\]
where
\[
-i{\Delta}_F(x-y) \equiv {\theta}(x^4-y^4) {\Delta}(x-y;m)
+ {\theta}(y^4-x^4) {\Delta}(x-y;-m),
\]
${\Delta}_F(x-y)$ being the Galilean Feynman propagator for a free
scalar field \cite{santos2}, and
\[
{\Delta}(x-y;\pm m) = \frac{1}{(2{\pi})^3}
\int d^3\bp e^{\mp \ri p (x-y)}.
\]

The positive- and negative-energy/mass contributions to ${\Delta}(x-y)$ can be
written explicitly as
\begin{equation}
\begin{array}{rcl}
{\Delta}(x-y)& = &\sqrt{2} \left[ e^{-\ri m\bar{c} (x^5-y^5)} G_{+}^0({\bf x} -
{\bf y}; x^4-y^4; m) \right. \\
& & \qquad\qquad  + \left. e^{\ri m\bar{c} (x^5-y^5)} G_{-}^0({\bf x} -
{\bf y}; x^4-y^4; -m) \right],
\label{deltaexpansion}
\end{array}
\end{equation}
where
\[
G_{+}^0({\bf x}-{\bf y};x^4-y^4;m) \equiv \frac{\ri}{(2{\pi})^3}
{\theta}(x^4-y^4) \int d^3\bp e^{\ri {\bf p}({\bf x}-{\bf y})
-\ri \frac{E}{\bar{c}} (x^4-y^4)}
\]
is the Schr\"{o}dinger Green's function of a scalar particle with mass
$m$ \cite{abreu}, and
\[
G_{-}^0({\bf x}-{\bf y};x^4-y^4;-m) =G_{+}^0({\bf y}-{\bf x};y^4-x^4;m).
\]
This ensures that Eq. (\ref{deltaexpansion}) is compatible with our
 earlier statements about the splitting of positive versus negative
 energy and mass.

With Eq. (\ref{deltaexpansion}), $S_1(x-y)$ becomes
\[
S_1(x-y)=\sqrt{2} S_F(x-y),
\]
where
\[
S_F(x-y) \equiv -\frac{1}{2k} (i{\gamma}^{\mu} {\partial}_{\mu} +k)
{\Delta}_F(x-y)
\]
is the Galilean Feynman propagator for a free Dirac field.
If we substitute Eqs. (\ref{deltatransform}) and (\ref{stransform}) into
Eq. (\ref{relation}), we find
\[
\bar{S}_1(p) = \frac{{\gamma}^{\mu} p_{\mu} + k}{p_{\mu}p^{\mu} - k^2
+ \ri{\varepsilon}}.
\]
Using Eq. (\ref{deltaexpansion}), we write $S_1(x-y)$ in the form :
\begin{equation}
S_1(x-y)=e^{-\ri m\bar{c} (x^5-y^5)} S_{1}(\bfx-\bfy, x^4-y^4; m)
+ e^{\ri m\bar{c} (x^5-y^5)} S_{1}(\bfx-\bfy, x^4-y^4; -m),
\label{sfactor}
\end{equation}
where
\[
S_1(\bfx-\bfy, x^4-y^4; \pm m)=\sqrt{2} S_{\pm}^0({\bf x} -{\bf y}; x^4-y^4;\pm m),
\]
and
\begin{eqnarray*}
S_{+}^0({\bf x}-{\bf y},x^4-y^4; m) & = &
\frac{1}{2k} {\gamma}^4 {\delta}(x^4-y^4) {\delta}({\bf x} -
{\bf y}) \\
& - &  {\theta}(x^4-y^4) \frac{\ri}{(2\pi)^3}
\int d^3\bp S_{+}({\bf p},m) e^{\ri [{\bf p} ({\bf x} -{\bf y}) -
\frac{E}{\bar{c}}(x^4-y^4)]}
\end{eqnarray*}
is the Schr\"{o}dinger Green's function of a Dirac particle of  mass $m$
with
\[
S_{+}({\bf p},m) = \frac{1}{4E_k}
\left( \begin{array}{cc}
E+3E_k & E-E_k - \ri \bar{c} \sqrt{2} {\bf\sigma}\cdot {\bf p}\\
-E+E_k - \ri \bar{c} \sqrt{2} {\bf\sigma}\cdot {\bf p} & -E+E_k
\end{array}
\right).
\]
For the negative-mass contribution, we have
\begin{eqnarray*}
S_{-}^0({\bf x}-{\bf y},x^4-y^4;- m) & = &
-\frac{1}{2k} {\gamma}^4 {\delta}(x^4-y^4) {\delta}({\bf x} -
{\bf y}) \\
& - &  {\theta}(y^4-x^4) \frac{\ri}{(2\pi)^3}
\int d^3\bp S_{-}({\bf p},m) e^{-\ri [{\bf p} ({\bf x} -{\bf y}) -
\frac{E}{\bar{c}}(x^4-y^4)]},
\end{eqnarray*}
where
\[
S_{-}({\bf p},m)=1 - S_{+}({\bf p},m).
\]

The propagator $S_F(x-y)$ can be defined in the canonical formalism as
well. The expansions in Eqs. (\ref{psiexpansion}) and
(\ref{barpsiexpansion}) yield an expression for the Feynman propagator as
\[
\langle 0|T\left[ {\Psi}(x) \bar{\Psi}(y) \right] |0 \rangle =
i S_F(x-y),
\]
where $T$ denotes the time ordering. This formula connects the path-integral
and canonical formalism and proves their equivalence.

Now let us define the Galilean one-particle Green's function for free fields :
\begin{equation}
G^{0}\left( x_1,x_2\right) =\left.\frac{\left( -\ri\right)
^{2}}{Z_{0}\left[ 0%
\right] }\frac{\delta ^{2}Z_{0}\left[ J,\barJ\right] }{\delta \barJ
\left( x_1\right) \delta J\left( x_2\right) }\right |_{_{J=0=\barJ}},
\label{FG 5dim}
\end{equation}
where $Z_{0}\left[ J,\barJ\right]$ is given in Eq. (\ref{z0final}),
 thus leading to
\begin{equation}
G^{0}\left( x_1,x_2\right) =-\ri \sqrt{2} S _{F}\left( x_1-x_2\right) .
\label{FG-PF}
\end{equation}
It is possible to calculate the average values of the
 translation generators in the Hilbert space given in
 Eq. (\ref{fivemomentum}), i.e. the observables corresponding to
 momentum, energy and mass in quantum mechanics :
\[
\left\langle \widehat{{\cal O}}\right\rangle =\ri\int
d^{3}x\ dx^{5}\lim_{y\rightarrow x}\left[ {\cal O}G^{0}\left( x,y\right) %
\right]
\]
where ${\cal O}$ denotes $P_i$, $H=P_4$ or $M=P_5$.

We may generalize the one-particle Green's function in Eq. (\ref{FG 5dim})
to the $n$-particle Green's functions :
\bey
G^{0}\left( x_{1},\ldots ,x_n ; y_{1},\ldots ,y_{n}\right)&=&
\langle 0|T(\Psi(x_1)\cdots\Psi(x_n)\barPsi(y_1)\cdots\barPsi(y_n)|0\rangle,\nonumber\\
&=&\frac{\left(-\ri\right) ^{2n}}{Z_{0}\left[ 0\right] }\left.\frac{\delta ^{2n}Z_{0}\left[
J,\barJ \right] }{\delta \barJ\left( x_{1}\right)\ldots \delta \barJ\left(
x_{n}\right) \delta J\left( y_{1}\right)\ldots \delta
J\left(y_{n}\right) }\right |_{_{J=0=\barJ}}.  \label{correlationf}
\eey
For instance, the 1-particle Green's function is given in Eq. (\ref{FG 5dim})
 and the 2-particle Green's function is given as
\bey
G^0(x_1,x_2;y_1,y_2)=\frac{1}{Z_0[0]}\left.\frac{\delta^4Z_0[J,\barJ]}
{\delta\barJ(x_1)\delta\barJ(x_2)\delta J(y_1)\delta J(y_2)}
\right|_{J=0=\barJ}.\label{4pcorrelationf}
\eey
Explicit forms and perturbative expansion will be given later on in an
 interacting system with quartic interactions.  For one-particle Green's
 function, an equation similar to the Schwinger-Dyson equation is obtained
 with the self-energy defined explicitly.

%%%%%%%%%%%%%%%%%%%%%%%%%%%%%%%%%%%%%%%%%%%%%%%%%%%%%%%%%%%%%%%%%%%%%
\section{Self-interacting quartic potential}
%%%%%%%%%%%%%%%%%%%%%%%%%%%%%%%%%%%%%%%%%%%%%%%%%%%%%%%%%%%%%%%%%%%%%

Now consider a Lagrangian which contains a non-trivial interacting potential :
\[
{\cal L}={\cal L}_{0}+{\cal L}_{{\rm int}},
\]
where ${\cal L}_{0}$ is given by Eq. (\ref{lagdirac}) and
 ${\cal L}_{{\rm int}}$ is the interaction term that
 depends on $\Psi$ and $\barPsi$.
With an arbitrary interaction, the generating functional is
\[
Z\left[ J,\barJ\right] =\frac{\int {\cal D}\Psi \int {\cal D}\barPsi
 \exp \left\{ \ri I-\ri\int d^{5}x\left[ J\left( x\right) \barPsi  \left(
x\right) +\barJ \left( x\right) \Psi \left( x\right) \right] \right\} }
{ \int {\cal D}\Psi \int {\cal D}\barPsi \exp \left( \ri I\right) },
\]
with $I$ given in Eq. (\ref{acao}), and where ${\cal L}_0$ is replaced by ${\cal L}$.
 Following standard methods \cite{fieldtheorybooks},
 we write the generating functional as
\begin{equation}
Z\left[ J,\barJ \right] =N\exp \left\{ -\ri\int d^{5}x\ {\cal L}_{{\rm int}}
\left[ \frac{1}{\ri}\frac{\delta }{\delta \barJ},\frac{1}{\ri}\frac{\delta }{
\delta J}\right] \right\} Z_{0}\left[ J,\barJ \right],  \label{inter}
\end{equation}
where $N$ is a normalization factor, and $Z_0$ is given by Eq. (\ref{z0final}).
 We derive the Green's functions from Eq. (\ref{correlationf})
 by replacing $Z_{0}\left[ J,\barJ\right]$
with $Z\left[ J,\barJ\right]$.

Consider an interaction Lagrangian in the form
\[
{\cal L}_{int}=g (\bar{\Psi}(x) {\Psi}(x) )^2.
\]
When we expand Eq. (\ref{inter}) in powers of $g$, then $Z\left[ J,\barJ\right]$ becomes
\[\begin{array}{l}
 Z\left[ J,\barJ \right] =N\exp \left\{ -\ri g\int d^{5}x\
 \left[ \frac{\delta}{\delta J(x)} \frac{\delta}{\delta\barJ(x)}
 \frac{\delta}{\delta J(x)}\frac{\delta}{\delta\barJ(x)} \right] \right\} Z_{0}\left[ J,\barJ \right]\\
\qquad =N\left\{1-\ri g\int
d^5z\ \left( \frac{\delta}{\delta J(z)} \frac{\delta}{\delta\barJ(z)}
 \frac{\delta}{\delta J(z)}\frac{\delta}{\delta\barJ(z)} \right) \right. \\
\qquad\qquad - \frac{g^2}{2} \left. \int d^5z d^5w\ \left(\frac{\delta}{\delta J(w)}
\frac{\delta}{\delta\barJ(w)}
 \frac{\delta}{\delta J(w)}\frac{\delta}{\delta\barJ(w)} \right)\
  \left(\frac{\delta}{\delta J(z)} \frac{\delta}{\delta\barJ(z)}
 \frac{\delta}{\delta J(z)}\frac{\delta}{\delta\barJ(z)} \right) +\right. \\
\qquad\qquad \left.+\ O\left( g^{3}\right) \right\}\ Z_{0}\left[ J,\barJ\right] .
\end{array}
\]
To zeroth order in $g$, we simply retrieve $Z_{0}\left[ J,\barJ\right]$.

Henceforth, we shall utilize the short-hand notation
\[
\langle \barJ_{\alpha}(x) S_{{\alpha}{\beta}}(x-y)
J_{\beta}(y)\rangle\equiv
 \int d^5x\; d^5y\ \barJ_{\alpha}(x) S_{{\alpha}{\beta}}(x-y)
J_{\beta}(y),
\]
where $S_{{\alpha}{\beta}}(x-y) \equiv (S_1)_{{\alpha}{\beta}}(x-y)$,
and ${\alpha},{\beta}=1,2,3,4$ indicate the component structure of
virtual sources. Thus we have
\[
Z_0 =
e^{-\ri\langle{\barJ}_{\alpha}(x)S_{{\alpha}{\beta}}(x-y)
J_{\beta}(y) \rangle}.
\]

To second order in $g$, we find
\begin{equation}
Z[J,\bar{J}]=N \left \{1-\ri g \int d^5z T(z) - \frac{g^2}{2} \int d^5z d^5w
T(z,w) \right \} Z_0[J,\bar{J}],
\label{secondorder}
\end{equation}
where
\[
T(z) \equiv  \frac{1}{Z_0[J,\bar{J}]}
\frac{\delta}{\delta J_{\alpha}(z)} \frac{\delta}{\delta\barJ_{\alpha}(z)}
 \frac{\delta}{\delta J_{\beta}(z)}\frac{\delta}{\delta\barJ_{\beta}(z)}
Z_0[J,\bar{J}] ={\rm  Tr} \left( S_1(0) \right)^2 - \left( {\rm Tr}S_1(0)
\right)^2
\]
\[
- 2{\ri} \left( S_{{\alpha}{\beta}}(0) - {\delta}_{{\alpha}{\beta}}
{\rm Tr}S_1(0) \right) \langle S_{{\beta}{\delta}}(z-v) J_{\delta}(v)
\rangle
\langle \bar{J}_{{\gamma}}(u) S_{{\gamma}{\alpha}}(u-z) \rangle
\]
\[
+ \langle S_{{\beta}{\delta}}(z-v) J_{\delta}(v) \rangle
\langle \bar{J}_{{\gamma}}(u) S_{{\gamma}{\beta}}(u-z) \rangle
\langle S_{{\alpha}\bar{\delta}}(z-w) J_{\bar{\delta}}(w) \rangle
\langle \bar{J}_{\bar{\gamma}}(y) S_{\bar{\gamma}{\alpha}}(y-z) \rangle,
\]
and
\[
T(z,w) \equiv \frac{1}{Z_0[J,\bar{J}]}
\frac{\delta}{\delta J_{\alpha}(w)} \frac{\delta}{\delta\barJ_{\alpha}(w)}
 \frac{\delta}{\delta J_{\beta}(w)}\frac{\delta}{\delta\barJ_{\beta}(w)}
\left( T(z) Z_0[J,\bar{J}] \right).
\]
The normalization factor in Eq. (\ref{secondorder}) is chosen in such way
that
$Z[0]=1$,
\[
N=Z_0^{-1}[0] \left( 1-\ri g \int d^5z T_0(z) - \frac{g^2}{2} \int d^5z d^5w
T_0(z,w) \right)^{-1},
\]
where
\[
T_0(z) \equiv T(z)|_{J=0=\bar{J}}, \qquad T_0(z,w) \equiv
T(z,w)|_{J=0=\bar{J}},
\]
excluding vacuum graphs from consideration.

%%%%%%%%%%%%%%%%%%%%%%%%%%%%%%%%%%%%%%%%%%%%%%%%%%%%%%%%%%%%%%%%%%%%
\subsection{2-point function}
%%%%%%%%%%%%%%%%%%%%%%%%%%%%%%%%%%%%%%%%%%%%%%%%%%%%%%%%%%%%%%%%%%%%

The 2-point function $G(x_1,x_2)$ is given by Eq. (\ref{FG 5dim}) with
$Z_0[J,\bar{J}]$ replaced by $Z[J,\bar{J}]$.
To zero-th order in $g$, $G(x_1,x_2)$ is clearly the same as for the
free field. To first order in $g$, we obtain
\begin{equation}
G_{{\alpha}{\beta}}(x_1,x_2)=G^0_{{\alpha}{\beta}}(x_1,x_2)-2g
\int d^5z G^0_{{\alpha}{\bar{\alpha}}}(x_1,z)
\left( S_{\bar{\alpha}\bar{\beta}}(0) - {\delta}_{\bar{\alpha}
\bar{\beta}} {\rm Tr}S_1(0) \right)
G^{0}_{\bar{\beta}{\beta}}(z,x_2).
\label{2point}
\end{equation}
Using the definition of $S_1(x-y)$ given in Eq. (\ref{stransform}),
yields the positive and negative-mass contributions to
$S_{\bar{\alpha}\bar{\beta}}(0)$,
\[
S_{\bar{\alpha}\bar{\beta}}(0;\pm m)=\frac{1}{(2{\pi})^3} \int
d^3\bp \int dp^5\;
\frac{\left[ \frac{1}{2mc} ({\gamma}^4 p^5 + k) \pm \frac{1}{2}
{\gamma}^5 \right]_{\bar{\alpha}\bar{\beta}}}{\pm p^5 -\frac{E}{c} +\ri
\frac{\varepsilon}{2m\bar{c}}}.
\]
Taking into account the identity,
\[
\frac{1}{{\omega} \pm i{\varepsilon}} = {\cal P}\frac{1}{{\omega}} \mp
\ri{\pi} \delta({\omega}),
\]
where ${\varepsilon}>0$ and ${\cal P}$ denotes the principal value,
non-diagonal elements of $S_{\bar{\alpha}\bar{\beta}}(0;m)$ and
$S_{\bar{\alpha}\bar{\beta}}(0;-m)$ can be shown to be equal in magnitude
and opposite in sign, so that they cancel each other, and
$S_{\bar{\alpha}\bar{\beta}}(0)=
S_{\bar{\alpha}\bar{\beta}}(0;m) + S_{\bar{\alpha}\bar{\beta}}(0;-m)$
is diagonal,
\begin{equation}
S_{\bar{\alpha}\bar{\beta}}(0)=\frac{1}{4} {\delta}_{\bar{\alpha}
\bar{\beta}} {\rm Tr}S_1(0),
\label{diagonal}
\end{equation}
\begin{center} \begin{picture}(300,100)(0,0)
\Line(0,50)(90,50)
\Text(0,39)[]{$x_1$}
\Text(90,39)[]{$x_2$}
\Text(145,50)[]{$+$}
\Line(200,50)(290,50)
\Text(200,39)[]{$x_1$}
\Text(290,39)[]{$x_2$}
\CArc(245,60)(10,0,360)
\Vertex(245,50){2}
\Text(245,39)[]{$z$}
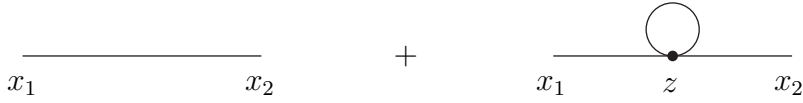
\end{picture} \\ {Figure 1: \footnotesize{\sl Diagrams for
the first order 2-point function for the quartic potential}}
\end{center}
where
\[
{\rm Tr}S_1(0) = 2{\rm Tr}S_1(0;m)=-\frac{\ri}{2\sqrt{2}{\pi}^3}
\int d^3\bp
\]
is a divergent quantity, which can be made finite with a cutoff.
 This serious divergence problem may be resolved by including a
 momentum-dependant vertex function. For example, if the vertex
 function decreases rapidly with momentum, then this integral
 can be convergent and finite.  Such a form may be anticipated
 for any realistic formulation of the problem.

With Eq. (\ref{diagonal}), the first order 2-point function can be rewritten as
\[
G_{{\alpha}{\beta}}(x_1,x_2)=G^{0}_{{\alpha}{\beta}}(x_1,x_2)
+ G^{(1)}_{{\alpha}{\beta}}(x_1,x_2),
\]
where
\[
G^{(1)}_{{\alpha}{\beta}}(x_1,x_2)
\equiv {\Sigma}_1(0)
\int d^5z G^{0}_{{\alpha}\bar{\alpha}}(x_1,z)
G^{0}_{\bar{\alpha}{\beta}}(z,x_2)
\]
and
\[
{\Sigma}_1(0) \equiv \frac{3g}{2} {\rm Tr}S_1(0),
\]
and represented diagrammatically in Fig. 1. The first diagram shows
the 2-point function for a free Galilean Dirac field. In the
second diagram, ${\Sigma}_1(0)$ is represented by a closed loop with one
vertex on it.

The second order in g contribution to the 2-point
function is
\[
G^{(2)}_{{\alpha}{\beta}}(x_1,x_2) =
\frac{g^2}{2} \left. \int d^5z d^5w
\frac{{\delta}^2}{{\delta}\bar{J}_{\alpha}(x_1)
{\delta}J_{\beta}(x_2)} \left( T(z,w) - 2 T_0(z) T(w) \right)
\right|_{J=0=\bar{J}},
\]
that is the sum of three terms
\begin{eqnarray}
G^{(2)}_{{\alpha}{\beta}}(x_1,x_2) & = &
g^2 \int d^5z d^5w
\left[ 4 G^{0}_{{\alpha}{\delta}}(x_1,w)
G^{0}_{{\delta}\bar{\alpha}}(w,z) \right.
{\Gamma}_{\bar{\alpha}\bar{\beta}}(z,w)
G^{0}_{\bar{\beta}{\beta}}(z,x_2)
\nonumber \\
& + & 3\ri {\rm Tr}S_1(0) \cdot G^{0}_{{\alpha}\bar{\alpha}}(x_1,z)
{\Gamma}_{\bar{\alpha}\bar{\beta}}(z,w) G^{0}_{\bar{\beta}{\beta}}(z,x_2)
\nonumber \\
& + & \frac{9}{4} {\left( {\rm Tr}S_1(0) \right)}^2 \cdot
G^{0}_{{\alpha}\bar{\alpha}}(x_1,z) \left.
G^{0}_{\bar{\alpha}\bar{\beta}}(z,w) G^{0}_{\bar{\beta}{\beta}}(w,x_2)
\right] .
\label{threeterms}
\end{eqnarray}
The corresponding diagrams are shown in Fig. 2. The function
${\Gamma}_{\bar{\alpha}\bar{\beta}}(z,w)$ in
Eq. (\ref{threeterms}) is defined as
\begin{equation}
{\Gamma}_{\bar{\alpha}\bar{\beta}}(z,w) \equiv
- G^0_{\bar{\alpha}{\delta}}(z,w) G^0_{{\delta}\bar{\beta}}(w,z)
+ {\delta}_{\bar{\alpha}\bar{\beta}} G^0_{{\gamma}{\delta}}(z,w)
G^0_{{\delta}{\gamma}}(w,z),
\label{gammazw}
\end{equation}
\begin{center} \begin{picture}(300,130)(0,0)
\Line(10,65)(90,65)
\Text(18,54)[]{$x_1$}
\Text(83,54)[]{$x_2$}
\CArc(50,65)(10,0,180)
\Vertex(40,65){2}
\CArc(50,65)(10,180,360)
\Vertex(60,65){2}
\Text(40,54)[]{$w$}
\Text(60,54)[]{$z$}
\Text(100,65)[]{$+$}
\Line(110,65)(190,65)
\Text(118,54)[]{$x_1$}
\Text(183,54)[]{$x_2$}
\Vertex(150,65){2}
\CArc(150,75)(10,0,360)
\Text(150,54)[]{$z$}
\Vertex(150,85){2}
\CArc(150,95)(10,0,360)
\Text(165,85)[]{$w$}
\Text(200,65)[]{$+$}
\Line(210,65)(290,65)
\Text(217,54)[]{$x_1$}
\Text(283,54)[]{$x_2$}
\Vertex(235,65){2}
\CArc(235,75)(10,0,360)
\Text(235,54)[]{$w$}
\Vertex(265,65){2}
\CArc(265,75)(10,0,360)
\Text(265,54)[]{$z$}
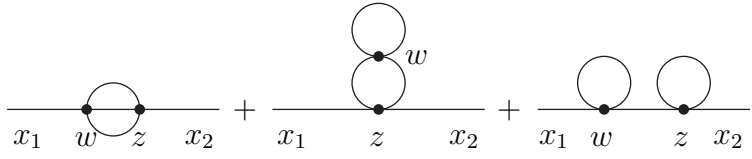
\end{picture} \\ {Figure 2: \footnotesize{\sl Diagrams for the $g^2$-order
of the 2-point function for the quartic potential. The third diagram is
not one-particle irreducible. The contribution of second diagram
vanishes.}} \end{center}
being represented by a closed loop with two vertices. Each of the vertices
can have up to two external lines. However, for the first diagram in
Fig. 2,
one leg of the vertex at $w$ is joined to a leg of the vertex at $z$
creating
an internal line, so that we have one external line at each vertex. At the
second diagram, two legs of the vertex at $w$ are joined together, producing
${\Sigma}_1(0)$. From Eq. (\ref{gammazw}), we deduce the following relations
for
${\Gamma}_{\bar{\alpha}\bar{\beta}}(z,w)$:

\[
{\Gamma}_{\bar{\alpha}\bar{\alpha}}(z,w) =
3 G^0_{{\gamma}{\delta}}(z,w) G^0_{{\delta}{\gamma}}(w,z),
\]
and
\[
{\Gamma}_{\bar{\alpha}\bar{\beta}}(z,z)= -\frac{3}{16} {\delta}_{\bar{\alpha}\bar{\beta}} \left( {\rm Tr}S_1(0)
\right)^2.
\]

The second diagram in Fig. 1 and the first two in Fig. 2 are
one-particle irreducible; they cannot be disconnected by cutting through
any one internal line. The third diagram in Fig. 2 is a chain of two
first-order one-particle irreducible graphs.

To simplify the expression for $G^{(2)}_{{\alpha}{\beta}}(x_1,x_2)$,
 Eq. (\ref{threeterms}), we can factor the fifth
coordinate out of the functions $G^0$ and ${\Gamma}$ and then perform
integrations over $z^5$ and $w^5$. The function
${\Gamma}_{\bar{\alpha}\bar{\beta}}(z,w)$ is factorized as follows
\begin{equation}
\begin{array}{rcl}
{\Gamma}_{\bar{\alpha}\bar{\beta}}(z,w)&=&
{\Gamma}^0_{\bar{\alpha}\bar{\beta}}({\bf z}, {\bf w}, z^4, w^4; m) +
e^{-2\ri m\bar{c}(z^5-w^5)} {\Gamma}^{(+)}_{\bar{\alpha}\bar{\beta}}
({\bf z}, {\bf w}, z^4, w^4; m) +\\
& &\qquad
+e^{2\ri m\bar{c}(z^5-w^5)} {\Gamma}^{(-)}_{\bar{\alpha}\bar{\beta}}
({\bf z}, {\bf w}, z^4, w^4; m),\end{array}
\label{gammafactor}
\end{equation}
where
\begin{equation}
\begin{array}{rcl}
{\Gamma}^0_{\bar{\alpha}\bar{\beta}}({\bf z}, {\bf w}, z^4, w^4; m)& \equiv &
-\left[ G^0_{\bar{\alpha}{\delta}}({\bf z}, {\bf w}, z^4, w^4; m)
 G^0_{{\delta}\bar{\beta}}({\bf w}, {\bf z}, w^4, z^4; m) +\right.\\
& & \qquad\quad \left. + G^0_{\bar{\alpha}{\delta}}
({\bf z}, {\bf w}, z^4, w^4; -m) G^0_{{\delta}\bar{\beta}}
 ({\bf w}, {\bf z}, w^4, z^4; -m) \right]+\\
& & \quad + {\delta}_{\bar{\alpha}\bar{\beta}}
\left[ G^0_{{\gamma}{\delta}}
({\bf z}, {\bf w}, z^4, w^4; m) G^0_{{\delta}{\gamma}}
({\bf w}, {\bf z}, w^4, z^4; m) +\right.\\
& &\qquad\left. + G^0_{{\gamma}{\delta}}
({\bf z}, {\bf w}, z^4, w^4; -m) G^0_{{\delta}{\gamma}}
({\bf w}, {\bf z}, w^4, z^4; -m) \right]\end{array}
\label{gammazero}
\end{equation}
is that part of ${\Gamma}_{\bar{\alpha}\bar{\beta}}(z,w)$,
which does not oscillate in the fifth coordinates, while
\[
\begin{array}{rcl}
{\Gamma}^{(+)}_{\bar{\alpha}\bar{\beta}}({\bf z}, {\bf w}, z^4, w^4; m)& \equiv &
- G^0_{\bar{\alpha}{\delta}}({\bf z}, {\bf w}, z^4, w^4; m) G^0_{{\delta}\bar{\beta}}
({\bf w}, {\bf z}, w^4, z^4; -m) +\\
& & \qquad + {\delta}_{\bar{\alpha}\bar{\beta}} G^0_{{\gamma}{\delta}}
({\bf z}, {\bf w}, z^4, w^4; m)
G^0_{{\delta}{\gamma}}({\bf w}, {\bf z}, w^4, z^4; -m)
\end{array}
\]
and
\[
{\Gamma}^{(-)}_{\bar{\alpha}\bar{\beta}}({\bf z}, {\bf w}, z^4, w^4; m) \equiv
{\Gamma}^{(+)}_{\bar{\alpha}\bar{\beta}}({\bf z}, {\bf w}, z^4, w^4; -m).
\]
Integrating both parts of Eq. (\ref{gammafactor}) over $w^5$ (or $z^5$) and
using the limit
\[
{\lim}_{l \to \infty} \frac{1}{l} \int_{-l/2}^{l/2} dw^5
e^{\pm 2\ri m\bar{c} w^5}= {\lim}_{l \to \infty} \frac{\sin(2m\bar{c}l)}{m\bar{c}l}=0
\]
valid for nonzero values of $m$, we obtain
\begin{equation}
\int dw^5 {\Gamma}_{\bar{\alpha}\bar{\beta}}(z,w)=
{\Gamma}^0_{\bar{\alpha}\bar{\beta}}({\bf z}, {\bf w}, z^4, w^4; m),
\label{gammaint}
\end{equation}
i.e. the oscillating parts of ${\Gamma}_{\bar{\alpha}\bar{\beta}}(z,w)$ do
not contribute to the integral.

Using the expressions given by Eqs. (\ref{sfactor}) and
(\ref{gammafactor}), we find that the product
\[
G^{0}_{{\alpha}{\delta}}(x_1,w)
G^{0}_{{\delta}\bar{\alpha}}(w,z)
{\Gamma}_{\bar{\alpha}\bar{\beta}}(z,w)
G^{0}_{\bar{\beta}{\beta}}(z,x_2)
\]
in the first term of the right-hand side of Eq. (\ref{threeterms})
has the following non-oscillating parts in $z^5$ and $w^5$:
\[
\begin{array}{l}
e^{-{\ri}m\bar{c}(x_1^5-x_2^5)}
G^{0}_{{\alpha}{\delta}}({\bf x}_1, {\bf w}, x_1^4, w^4; m)
G^{0}_{{\delta}\bar{\alpha}}({\bf w}, {\bf z}, w^4, z^4; m)\times\\
\qquad\times{\Gamma}^0_{\bar{\alpha}\bar{\beta}}({\bf z}, {\bf w}, z^4, w^4; m)
G^{0}_{\bar{\beta}{\beta}}({\bf z}, {\bf x}_2, z^4, x_2^4; m)+\\
+ e^{{\ri}m\bar{c}(x_1^5-x_2^5)}
G^{0}_{{\alpha}{\delta}}({\bf x}_1, {\bf w}, x_1^4, w^4; -m)
G^{0}_{{\delta}\bar{\alpha}}({\bf w}, {\bf z}, w^4, z^4; -m)\times\\
\qquad\times{\Gamma}^0_{\bar{\alpha}\bar{\beta}}({\bf z}, {\bf w}, z^4, w^4; -m)
G^{0}_{\bar{\beta}{\beta}}({\bf z}, {\bf x}_2, z^4, x_2^4; -m),
\end{array}
\]
so that the corresponding integral can be written as sum of positive- and
negative-mass contributions. Let us represent
$G^{0}_{{\alpha}{\delta}}({\bf x}_1, {\bf w}, x_1^4, w^4; m)$
by a line in the $(3+1)$ space-time, with an arrow pointed in the
direction in which the particle is moving, i.e. from $x_1$ to $w$, and
$G^{0}_{{\alpha}{\delta}}({\bf x}_1, {\bf w}, x_1^4, w^4; -m)$
by a line again running from $x_1$ to $w$ and carrying an arrow in the
opposite direction, as in Fig. 3.

%\Text(0,40)[l]{$G^0_{\alpha\delta}({\bf x}_1,{\bf w},x_1^4, w^4; m):$}
%\ArrowLine(120,40)(160,40)
%\Text(120,30)[]{$(\bfx_1, x_1^4)$}
%\Text(160,30)[]{$(\bfw, w^4)$}
%\Text(190,40)[l]{$; G^0_{\alpha\delta}({\bf x}_1, {\bf w}, x_1^4, w^4; -m):$}
%\ArrowLine(360,40)(320,40)
%\Text(320,30)[]{$(\bfx_1, x_1^4)$}
%\Text(360,30)[]{$(\bfw, w^4)$}

\begin{center}\begin{picture}(330,160)(0,0)
\Text(0,120)[l]{$(a)\qquad G^0_{\alpha\delta}({\bf x}_1,{\bf w},x_1^4, w^4; m)\ :$}
\ArrowLine(200,120)(300,120)
\Text(200,110)[]{$(\bfx_1, x_1^4)$}
\Text(300,110)[]{$(\bfw, w^4)$}
\Text(0,40)[l]{$(b)\qquad G^0_{\alpha\delta}({\bf x}_1,{\bf w},x_1^4, w^4; -m)\ :$}
\ArrowLine(300,40)(200,40)
\Text(200,30)[]{$(\bfx_1, x_1^4)$}
\Text(300,30)[]{$(\bfw, w^4)$}
\end{picture} \\ {Figure 3: \footnotesize{\sl
Green's functions in $(3+1)$ dimensions for (a) positive and
 (b) negative energy/mass}}
\end{center}

When the Green's function $G^0_{\alpha\beta}$, defined
 in $(4+1)$-dimensions, is reduced to $(3+1)$ Galilean space-time,
 it contains two parts, one for positive energy
 and the other for negative energy. If we represent the
 $(4+1)$-dimensional Green's function by a simple line, and the
 $(3+1)$-dimensional Green's functions by a line containing an arrow,
 then the relation between the $(4+1)$-dimensional Green's function
 and the positive- and negative-energy/mass contributions in $(3+1)$ space-time
 diagrams is represented as in Fig. 4. The symmetry of the two-point Green's
 function under interchange of $x_1$ and $w$ is obvious.

\begin{center} \begin{picture}(350,160)(0,0)
\Text(70,120)[r]{$G^0_{\alpha\beta}(x_1,w)\ = $}
\Line(90,120)(190,120)
\Text(90,110)[]{$x_1$}
\Text(190,110)[]{$w$}
\Text(40,70)[l]{$=e^{-\ri m(x_1^5-w^5)}$}
\ArrowLine(120,70)(180,70)
\Text(120,60)[]{$(\bfx_1, x_1^4)$}
\Text(180,60)[]{$(\bfw, w^4)$}
\Text(205,70)[l]{$+\ e^{+\ri m(x_1^5-w^5)}$}
\ArrowLine(345,70)(285,70)
\Text(285,60)[]{$(\bfx_1, x_1^4)$}
\Text(345,60)[]{$(\bfw, w^4)$}
\end{picture} \\ {Figure 4: \footnotesize{\sl Positive- and
 negative-energy/mass contributions to the Green's function through
 dimensional reduction}}
\end{center}

For instance, the total contribution of the first
 one-particle-irreducible second-order diagram
 (in $(4+1)$ space-time) in Fig. 2 contains four diagrams
 after reduction to $(3+1)$ space-time. These diagrams
 are shown in Fig. 5.  Although the number of
diagrams increases after the reduction to $(3+1)$-dimensions, a clear
interpretation in terms of particles and antiparticles becomes
possible.  The first two diagrams with external lines being particles
represent the second-order positive-mass contribution to the $2$-point
function, while two others with external lines being antiparticles
represent the second-order negative-mass contribution.

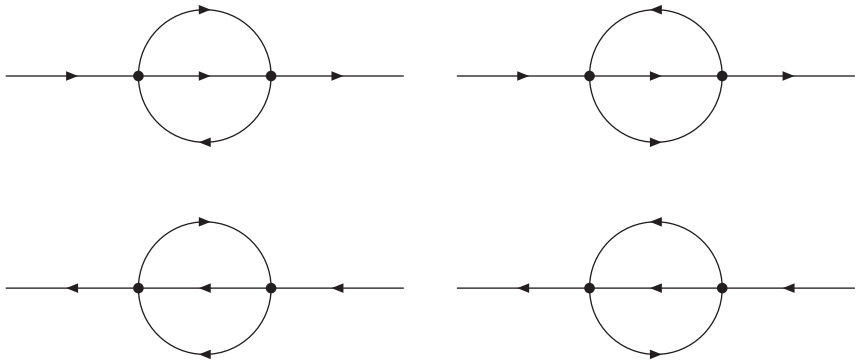
\begin{center}\begin{picture}(330,160)(0,0)
\ArrowLine(0,120)(50,120)
\Vertex(50,120){2}
\ArrowLine(50,120)(100,120)
\Vertex(100,120){2}
\ArrowLine(100,120)(150,120)
\ArrowArcn(75,120)(25,180,0)
\ArrowArcn(75,120)(25,360,180)
\ArrowLine(170,120)(220,120)
\Vertex(220,120){2}
\ArrowLine(220,120)(270,120)
\Vertex(270,120){2}
\ArrowLine(270,120)(320,120)
\ArrowArc(245,120)(25,0,180)
\ArrowArc(245,120)(25,180,360)
\ArrowLine(50,40)(0,40)
\Vertex(50,40){2}
\ArrowLine(100,40)(50,40)
\Vertex(100,40){2}
\ArrowLine(150,40)(100,40)
\ArrowArcn(75,40)(25,180,0)
\ArrowArcn(75,40)(25,360,180)
\ArrowLine(220,40)(170,40)
\Vertex(220,40){2}
\ArrowLine(270,40)(220,40)
\Vertex(270,40){2}
\ArrowLine(320,40)(270,40)
\ArrowArc(245,40)(25,0,180)
\ArrowArc(245,40)(25,180,360)
\end{picture} \\ {Figure 5: \footnotesize{\sl
 Contributions in $(3+1)$ space-time
 of the leftmost diagram in Fig. 2}}
\end{center}

Performing the integration over $w^5$ in the second term
of the right-hand side of Eq. (\ref{threeterms})
as well and using Eq. (\ref{gammaint}), this gives
\[
\int d^5w {\Gamma}_{\bar{\alpha}\bar{\beta}}(z,w)=
\int d^3{\bf w} \int dw^4 {\Gamma}^0_{\bar{\alpha}\bar{\beta}}
 ({\bf z}, {\bf w}, z^4, w^4; m).
\]
It is clear from the expression for
${\Gamma}^0_{\bar{\alpha}\bar{\beta}}({\bf z}, {\bf w}, z^4, w^4; m)$,
Eq. (\ref{gammazero}), that the positive- and negative-
mass Green's functions contribute separately. Calculating these contributions, we have
\[
\int d^3{\bf w} \int dw^4
G^0_{\bar{\alpha}{\delta}}({\bf z}, {\bf w}, z^4, w^4; \pm m)
 G^0_{{\delta}\bar{\beta}}({\bf w}, {\bf z}, w^4, z^4; \pm m)
= \mp \frac{1}{8m\bar{c}} ({\gamma}^4)_{\bar{\alpha}\bar{\beta}}
{\rm Tr}S_1(0),
\]
that results in
\[
\int d^3{\bf w} \int dw^4 {\Gamma}^0_{\bar{\alpha}\bar{\beta}}({\bf z}, {\bf w}, z^4, w^4; m)=0.
\]
Therefore, the second diagram in Fig. 2 does not contribute.

Let us write the sum of one-particle-irreducible graphs for $2$-point function in all orders in $g$ as
\[
\int d^5z d^5w G^{0}_{{\alpha}\bar{\alpha}}(x_1,z)
{\Sigma}_{\bar{\alpha}\bar{\beta}}(z,w)
G^{0}_{\bar{\beta}{\beta}}(w,x_2),
\]
with two external lines and one self-energy subgraph
 ${\Sigma}_{\bar{\alpha}\bar{\beta}}(z,w)$. To order $g^2$, we have
\[
{\Sigma}_{\bar{\alpha}\bar{\beta}}(z,w)=
{\delta}_{\bar{\alpha}\bar{\beta}} {\delta}^{(5)}(z-w) {\Sigma}_1(0)
+ 4g^2 G^{0}_{\bar{\alpha}{\delta}}(w,z) {\Gamma}_{{\delta}\bar{\beta}}(z,w)
+ O(g^3).
\]
Then the complete $2$-point function is given by a sum of chains of one, two, and more of these subgraphs connected
with the free field propagators
\begin{equation}
\begin{array}{l}
G_{{\alpha}{\beta}}(x_1,x_2)  =
G^{0}_{{\alpha}{\beta}}(x_1,x_2) +
\int d^5z d^5w G^{0}_{{\alpha}\bar{\alpha}}(x_1,z)
{\Sigma}_{\bar{\alpha}\bar{\beta}}(z,w)
G^{0}_{\bar{\beta}{\beta}}(w,x_2) + \\
\qquad +  \int d^5z d^5w \int d^5\bar{z} d^5\bar{w} G^{0}_{{\alpha}\bar{\alpha}}(x_1,z)
{\Sigma}_{\bar{\alpha}\bar{\beta}}(z,w)
G^{0}_{\bar{\beta}{\delta}}(w,\bar{z})
{\Sigma}_{{\delta}\bar{\delta}}(\bar{z},\bar{w})
G^{0}_{\bar{\delta}{\beta}}(\bar{w},x_2)+\cdots \\
\qquad = G^{0}_{{\alpha}{\beta}}(x_1,x_2)
+ \int d^5z d^5w G^{0}_{{\alpha}\bar{\alpha}}(x_1,z)
{\Sigma}_{\bar{\alpha}\bar{\beta}}(z,w)
G_{\bar{\beta}{\beta}}(w,x_2).
\label{closedeq}
\end{array}
\end{equation}
Introducing the Fourier transform ${\Sigma}_{{\alpha}{\beta}}(p)$ and $G_{{\alpha}{\beta}}(p)$ of the
functions ${\Sigma}_{{\alpha}{\beta}}(x_1,x_2)$ and  $G_{{\alpha}{\beta}}(x_1,x_2)$ in the same way as the Fourier
transforms of the Galilean propagators in equations (\ref{deltatransform}) and
(\ref{stransform}), we rewrite Eq. (\ref{closedeq}) in $(4+1)$-dimensional momentum space
as
\[
{\ri} G_{{\alpha}{\beta}}(p) =
S_{{\alpha}{\beta}}(p)
+ S_{{\alpha}\bar{\alpha}}(p) {\Sigma}_{\bar{\alpha}\bar{\beta}}(p) G_{\bar{\beta}{\beta}}(p)
\]
that results in the exact expression for $G_{{\alpha}{\beta}}(p)$:
\[
G_{{\alpha}{\beta}}(p) =
\left[ {\ri}S_1^{-1}(p) - {\Sigma}(p) \right]^{-1}_{{\alpha}{\beta}}.
\]
This expression is similar to the case of many-body systems where the
 exact expression of the 2-point function depends on self-energy.

%%%%%%%%%%%%%%%%%%%%%%%%%%%%%%%%%%%%%%%%%%%%%%%%%%%%%%%%%%%%%%%%%%%%%%
\subsection{4-point function}
%%%%%%%%%%%%%%%%%%%%%%%%%%%%%%%%%%%%%%%%%%%%%%%%%%%%%%%%%%%%%%%%%%%%%%%%%

The 4-point function $G(x_1,x_2;y_1,y_2)$ is given by Eq. (\ref{4pcorrelationf})
 with $Z_0[J,\bar{J}]$ replaced by $Z[J,\bar{J}]$.
To find its irreducible part, we can use a generating functional
$W[J,\bar{J}]$, which generates only connected Feynman diagrams or
connected Green's functions. It is related to $Z[J,\bar{J}]$ as
\[
W[J,\bar{J}]= -\ri \ln Z[J,\bar{J}].
\]
We define the irreducible or connected 4-point function as
\begin{equation}
\bar{G}(x_1,x_2;y_1,y_2)=\frac{1}{W[0]}\left.\frac{\delta^4W[J,\barJ]}
{\delta\barJ(x_1)\delta\barJ(x_2)\delta J(y_1)\delta J(y_2)}
\right|_{J=0=\barJ},\label{irreducible}
\end{equation}
that gives us the following relation between $\bar{G}(x_1,x_2;y_1,y_2)$
and the complete 4-point function $G(x_1,x_2;y_1,y_2)$:
\begin{equation}
\begin{array}{rcl}
\bar{G}_{{\alpha}{\beta}{\gamma}{\delta}}(x_1,x_2;y_1,y_2)
& = & -\ri G_{{\alpha}{\beta}{\gamma}{\delta}}(x_1,x_2;y_1,y_2) + \\
& + & \ri \left[
G_{{\alpha}{\delta}}(x_1,y_2)
G_{{\beta}{\gamma}}(x_2,y_1) -
G_{{\alpha}{\gamma}}(x_1,y_1)
G_{{\beta}{\delta}}(x_2,y_2)
\right].
\end{array}
\label{barG59}
\end{equation}

To order $g^0$, the complete 4-point function contains only reducible parts,
\begin{equation}
G^0_{{\alpha}{\beta}{\gamma}{\delta}}(x_1,x_2;y_1,y_2)
= G^0_{{\alpha}{\delta}}(x_1,y_2)
G^0_{{\beta}{\gamma}}(x_2,y_1) -
G^0_{{\alpha}{\gamma}}(x_1,y_1)
G^0_{{\beta}{\delta}}(x_2,y_2).
\label{HF}\end{equation}
This represents the Hartree-Fock part of the 4-point function. All remaining
 parts include interaction among the particles. Using $\bar{G}^0$
 defined in Eq. (\ref{barG59}), we find that
 $\bar{G}^0_{{\alpha}{\beta}{\gamma}{\delta}}(x_1,x_2;y_1,y_2)=0$.
 The diagrams corresponding to Eq. (\ref{HF}) are shown in Fig. 6.

\begin{center} \begin{picture}(300,130)(0,0)
\Text(-50,65)[]{$-$}
\Line(0,25)(90,25)
\Text(5,14)[]{$x_2$}
\Text(90,14)[]{$y_2$}
\Text(45,5)[]{}
\Line(0,105)(90,105)
\Text(5,91)[]{$x_1$}
\Text(90,91)[]{$y_1$}
\Line(200,25)(245,65)
\Text(205,14)[]{$x_2$}
\Text(145,65)[]{$+$}
\Text(290,14)[]{$y_2$}
\Text(45,5)[]{}
\CArc(245,65)(10,-45,135)
\Line(200,105)(238,72)
\Text(205,91)[]{$x_1$}
\Line(245,65)(290,105)
\Line(252,58)(290,25)
\Text(290,91)[]{$y_1$}
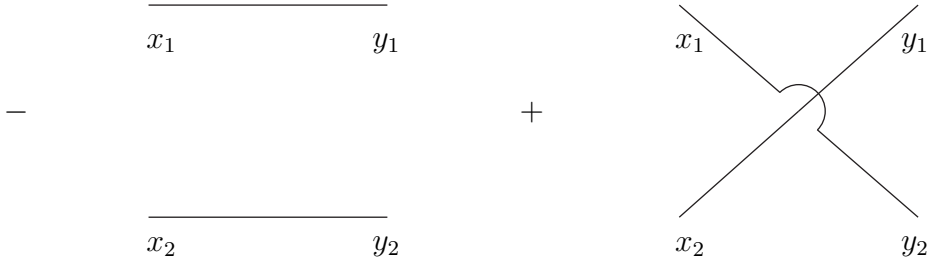
\end{picture} \\ {Figure 6: \footnotesize{\sl Diagrams for
$g^0$-order
of the 4-point function for the quartic
potential}}\end{center}

The irreducible parts appear in the  first order in $g$,
\[
\begin{array}{lcl}
\bar{G}^{(1)}_{{\alpha}{\beta}{\gamma}{\delta}}(x_1,x_2;y_1,y_2)
& = & -2g \int dz^5
\left[
G^0_{{\alpha}\bar{\alpha}}(x_1,z)
G^0_{\bar{\alpha}{\delta}}(z,y_2)
G^0_{{\beta}\bar{\beta}}(x_2,z)
G^0_{\bar{\beta}{\gamma}}(z,y_1) \right. \\
& - & ( ({\delta},y_2) \leftrightarrow \left. ({\gamma},y_1) ) \right],
\end{array}
\]
where
$({\delta},y_2) \leftrightarrow ({\gamma},y_1)$ means that there
is an additional term in the square brackets, which can be
obtained from the first one by replacing $({\delta},y_2)$ with
$({\gamma},y_1)$ and vice versa. These parts are represented
diagrammatically in Fig. 7.

\begin{center} \begin{picture}(300,130)(0,0)
\Line(0,25)(45,65)
\Text(5,14)[]{$x_2$}
\Line(0,105)(45,65)
\Text(5,91)[]{$x_1$}
\Vertex(45,65){2}
\Line(45,65)(90,105)
\Text(90,14)[]{$y_2$}
\Line(45,65)(90,25)
\Text(90,91)[]{$y_1$}
\Text(140,65)[l]{$-\qquad (y_1\leftrightarrow y_2)$}
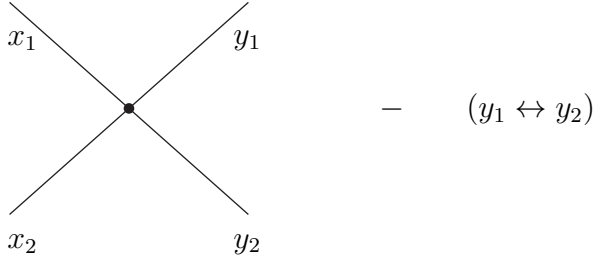
\end{picture} \\ { Figure 7: \footnotesize{\sl Diagrams for $g^1$-order
 of the irreducible 4-point Green's function for the quartic potential}}\end{center}

The second order in $g$ contribution to the irreducible 4-point
function is
\[
\begin{array}{l}
\bar{G}^{(2)}_{{\alpha}{\beta}{\gamma}{\delta}}(x_1,x_2;y_1,y_2)
= {\ri}
\left[
G^{(1)}_{{\beta}{\gamma}}(x_2,y_1)
G^{(1)}_{{\alpha}{\delta}}(x_1,y_2) -
G^{(1)}_{{\alpha}{\gamma}}(x_1,y_1)
G^{(1)}_{{\beta}{\delta}}(x_2,y_2) \right]+\\
+ {\ri}\; \frac{g^2}{2} \left. \int d^5z d^5w
\frac{{\delta}^4}{{\delta}\bar{J}_{\alpha}(x_1)
{\delta}\bar{J}_{\beta}(x_2) {\delta}J_{\gamma}(y_1)
{\delta}J_{\delta}(y_2)} \left( T(z,w) - 2
T_0(z) T(w) \right)
\right|_{J=0=\bar{J}},
\end{array}
\]
that can be rewritten as
\[
\begin{array}{l}
\bar{G}^{(2)}_{{\alpha}{\beta}{\gamma}{\delta}}(x_1,x_2;y_1,y_2)
 =  -2g \int d^5w \left[
G^0_{{\alpha}\bar{\alpha}}(x_1,w)
G^0_{\bar{\alpha}{\delta}}(w,y_2)
G^0_{{\beta}\bar{\beta}}(x_2,w)
G^{(1)}_{\bar{\beta}{\gamma}}(w,y_1) \right. + \\
\qquad\quad +
G^0_{{\alpha}\bar{\alpha}}(x_1,w)
G^0_{\bar{\alpha}{\delta}}(w,y_2)
G^{(1)}_{{\beta}\bar{\beta}}(x_2,w)
G^0_{\bar{\beta}{\gamma}}(w,y_1) + \\
\qquad\quad +
G^0_{{\alpha}\bar{\alpha}}(x_1,w)
G^{(1)}_{\bar{\alpha}{\delta}}(w,y_2)
G^0_{{\beta}\bar{\beta}}(x_2,w)
G^0_{\bar{\beta}{\gamma}}(w,y_1) + \\
\qquad\quad +
\left.
G^{(1)}_{{\alpha}\bar{\alpha}}(x_1,w)
G^0_{\bar{\alpha}{\delta}}(w,y_2)
G^0_{{\beta}\bar{\beta}}(x_2,w)
G^0_{\bar{\beta}{\gamma}}(w,y_1)
\right] \\
\quad -  4{\ri} g^2 \int d^5z d^5w
\left[
G^0_{{\alpha}\bar{\alpha}}(x_1,w)
G^0_{{\beta}\bar{\beta}}(x_2,z)
{\Gamma}_{\bar{\beta}\bar{\gamma},\bar{\alpha}\bar{\delta}}(z,w)
G^0_{\bar{\gamma}{\gamma}}(z,y_1)
G^0_{\bar{\delta}{\delta}}(w,y_2) \right. \\
\quad  +  \left.
G^0_{{\alpha}\bar{\alpha}}(x_1,w)
G^0_{\bar{\alpha}\bar{\delta}}(w,z)
G^0_{\bar{\delta}{\delta}}(z,y_2)
G^0_{{\beta}\bar{\beta}}(x_2,w)
G^0_{\bar{\beta}\bar{\gamma}}(w,z)
G^0_{\bar{\gamma}{\gamma}}(z,y_1) \right]\\
\qquad\qquad - ( ({\delta},y_2) \leftrightarrow  ({\gamma},y_1) ).
\end{array}
\]
Some diagrams representing these processes are shown in
 Fig. 8. Part (a) corresponds to the third line of the
 previous equation; there are three more similar diagrams
 with self-energy loop $\Sigma_1(0)$ on one of the remaining
 three legs.  Part (b) represents line 6, and
 part (c) corresponds to line 5 of the equation above.

\begin{center} \begin{picture}(360,130)(0,0)
\Line(30,25)(60,65)
\Text(20,25)[]{$x_2$}
\Line(30,105)(60,65)
\Text(20,105)[]{$x_1$}
\Vertex(60,65){2}
\Text(60,50)[]{$w$}
\Line(90,25)(60,65)
\Text(100,25)[]{$y_2$}
\Vertex(75,45){2}
\CArc(85,55)(14,0,360)
\Text(75,30)[]{$z$}
\Line(90,105)(60,65)
\Text(100,105)[]{$y_1$}
\Text(60,5)[]{$(a)$}
\Text(120,65)[]{$;$}
\Line(150,25)(165,65)
\Text(140,25)[]{$x_2$}
\Line(150,105)(165,65)
\Text(140,105)[]{$x_1$}
\Vertex(165,65){2}
\Text(165,45)[]{$w$}
\CArc(180,65)(15,0,360)
\Text(195,45)[]{$z$}
\Vertex(195,65){2}
\Line(195,65)(210,105)
\Text(220,105)[]{$y_1$}
\Line(195,65)(210,25)
\Text(220,25)[]{$y_2$}
\Text(180,5)[]{$(b)$}
\Text(240,65)[]{$;$}
\Line(270,25)(330,25)
\Text(260,25)[]{$x_2$}
\Text(300,15)[]{$z$}
\Text(300,5)[]{$(c)$}
\Text(340,25)[]{$y_1$}
\Vertex(300,25){2}
\CArc(250,65)(65,-40,40)
\CArc(350,65)(65,140,220)
\Line(270,105)(330,105)
\Text(260,105)[]{$x_1$}
\Text(300,115)[]{$w$}
\Text(340,105)[]{$y_2$}
\Vertex(300,105){2}
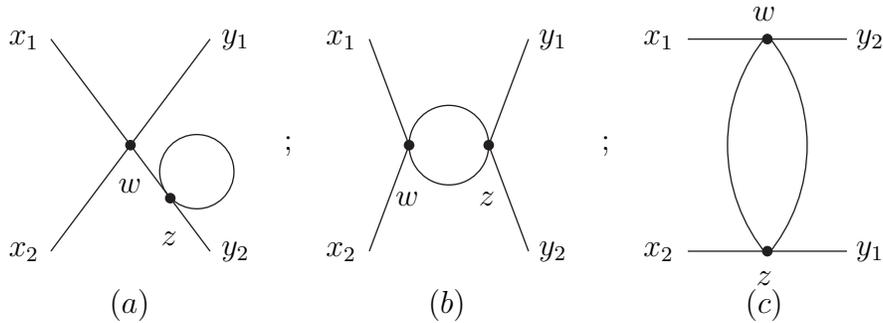
\end{picture} \\ { Figure 8: \footnotesize{\sl Diagrams for $g^2$-order
 of the irreducible 4-point Green's function for the quartic potential}}\end{center}

The function
${\Gamma}_{\bar{\beta}\bar{\gamma},\bar{\alpha}\bar{\delta}}(z,w)$
is defined as
\begin{eqnarray}
{\Gamma}_{\bar{\beta}\bar{\gamma},\bar{\alpha}\bar{\delta}}(z,w) & \equiv
& G^0_{\bar{\beta}\bar{\delta}}(z,w) G^0_{\bar{\alpha}\bar{\gamma}}(w,z)
+ {\delta}_{\bar{\alpha}\bar{\delta}} {\Gamma}_{\bar{\beta}\bar{\gamma}}
(z,w)
\nonumber \\
& + & {\delta}_{\bar{\beta}\bar{\gamma}}
{\Gamma}_{\bar{\alpha}\bar{\delta}}(w,z)
- 3 {\delta}_{\bar{\alpha}\bar{\delta}}
{\delta}_{\bar{\beta}\bar{\gamma}}
G^0_{{\gamma}{\delta}}(z,w)
G^0_{{\delta}{\gamma}}(w,z),
\end{eqnarray}
being represented by a closed loop with two vertices and four external
lines, as in Fig. 8 (c). This function has the following symmetry property
\[
{\Gamma}_{\bar{\beta}\bar{\gamma},\bar{\alpha}\bar{\delta}}(z,w)
={\Gamma}_{\bar{\alpha}\bar{\delta},\bar{\beta}\bar{\gamma}}(w,z).
\]
Taking $\bar{\beta}=\bar{\gamma}$ and summing over $\bar{\beta}$, this
gives us
\[
{\Gamma}_{\bar{\beta}\bar{\beta},\bar{\alpha}\bar{\delta}}(z,w)
=3 {\Gamma}_{\bar{\alpha}\bar{\delta}}(w,z)
-8 {\delta}_{\bar{\alpha}\bar{\delta}}
G^0_{{\gamma}{\delta}}(z,w)
G^0_{{\delta}{\gamma}}(w,z).
\]
In a similar way, we obtain
\[
{\Gamma}_{\bar{\beta}\bar{\gamma},\bar{\alpha}\bar{\alpha}}(z,w)
=3 {\Gamma}_{\bar{\beta}\bar{\gamma}}(z,w)
-8 {\delta}_{\bar{\beta}\bar{\gamma}}
G^0_{{\gamma}{\delta}}(z,w)
G^0_{{\delta}{\gamma}}(w,z).
\]

The polarisation part arise from the 4-point function and provides a sum
 of the loops to arbitrary order.  The equivalence to the case of
 non-relativistic many-body systems interacting by two-particle
 interactions is obvious.  Here we have parts that may be considered
 for particles and anti-particles.

\section{Concluding remarks}

This paper is the continuation of our previous works on quantization of
 Galilean-covariant field theories: path-integral quantization of complex
 scalar fields in Ref.  \cite{abreu} and the canonical quantization of both
 scalar and fermi fields in Ref. \cite{santos2}. The main purpose of this
 approach is to exploit relativistic tensorial techniques for applications
 to non-relativistic many-body systems. It is also interesting to
 compare Lorentzian and Galilean theories. An example of
 a rather unexpected similarity is that the non-zero spin is also
 predicted within a Galilean framework coherently
 defined \cite{levyleblond, levyleblond1967}.
 The presence of antiparticles is another example.  However, there is
 no creation of particle and antiparticle pairs.  It may be emphasized
 that the antisymmetrisation of the 4-point functions for fermions
 is also clearly respected. In addition to many
 familiar dissimilarities, some deserve to be emphasized, such as
 the existence of {\em two} Galilean formulations of electrodynamics
 \cite{bellac,crawford}.

We have discussed the Dirac equation on a $(4+1)$ manifold
 and its reduction to the L\'evy-Leblond equations
 \cite{levyleblond1967}, and the coexistence of positive- and
 negative-energy/mass solutions \cite{horzela}. While doing so,
 a representation of the Dirac matrices different from
 what is used earlier is presented, as well as the related spinors. After
 discussing the Galilean generating functional and Green's
 functions for particles and antiparticles, we compute the
 2- and 4-point functions for the self-interacting quartic
 potential.

From this study, we find that the following observations on the
 use of a $(4+1)$-dimensional Galilean space-time are in order.
 There exists a mass superselection rule, which prevents the
 creation of massive particles. This makes
 Yukawa coupling to massive particles irrelevant, because
 only couplings that involve at least four particles are
 allowed whereas Yukawa coupling to massless particles like
 the photon is possible.

An open question concerns the parity operator in Galilean
 field theories.  The Clifford algebra theory asserts
 that there is no parity operator analogous to $\gamma^5$
 in  any odd-dimensions.  We are currently
 investigating the possibility to embed the Galilean
  space-time into a $(5+1)$-dimensional
 Minkowski manifold, for which a parity operator exists
 with natural 8-dimensional Dirac gamma matrices
 \cite{kobayashi}. This extention will allow us to study
 the Galilean analogue of the Nambu-Jona-Lasinio model
 \cite{njl}.

%Finally, let us comment on the definition of mass. It occurs
%in two independent ways:
%through the constant $k$ in  Eq. (\ref{lagdirac}), and
%as the momentum's fifth component in Eq. (\ref{fivemomentum}),
%that is, as the conjugate variable of the extra parameter $s$.
%When we needed to perform the renormalization of mass, we
%have used the first point of view. Indeed, the appearance
%of $k$ in Dirac equation is compatible with the Galilean
%covariant equation $p^\mu p_\mu=k^2$, which suggests that
%$k$ is the Galilean analogue of the relativistic mass.

In Poincar\'e covariant field theories in $(3+1)$ dimensions,
 $p_\mu p^\mu=E^2-\bp^2c^2$ is an invariant and is equal to
 $m^2c^4$. This mass is an invariant quantity. However, in
 Galilean covariant field theories in $(4+1)$ dimensions,
 we have that $p_\mu p^\mu = 2mE-\bp^2$ is an invariant
 that is set equal to a constant $k^2$. It is important to
 emphasize that $m$ appears as central charge in the Galilean
 algebra and this leads to the definition of the five-momentum
 as $(\bp, E/{\bar c}, m{\bar c})$. Thus the renormalization process
 would affect the invariant $k$ in the Galilean covariant
 theory.

Finally, with the set up of the functional form for the
 path-integral approach, this would allow us to write down
 the transition amplitudes, hence the cross-sections, with the
 usual process of combining the square of the transition
 amplitude and the necessary phase space. The formulation
 as presented here has established contact with the usual
 perturbation theory for non-relativistic systems. However,
 it is important to emphasize that a covariant Galilean
 field theory is compatible with the idea of particles with
 energy $E$ and mass $+m$, and antiparticles with
 energy $-E$ and mass $-m$.

\section*{Acknowledgments}

We acknowledge partial support by the Natural Sciences and Engineering
Research  Council of Canada.

\end{document}